\documentclass{pas}
\usepackage{aas_macros}

\usepackage{natbib}
\usepackage{amsmath}	
\usepackage{wasysym}    
\usepackage{bm}         
\usepackage{xspace}
\usepackage{xcolor}

\newcommand\HI{\textsc{H\,i}\xspace} 
\newcommand\HII{\textsc{H\,ii}\xspace}

\newcommand\orcids[1]{\href{https://orcid.org/#1}{\includegraphics[height=8.5pt,trim={-6pt 0 -6pt 0},clip]{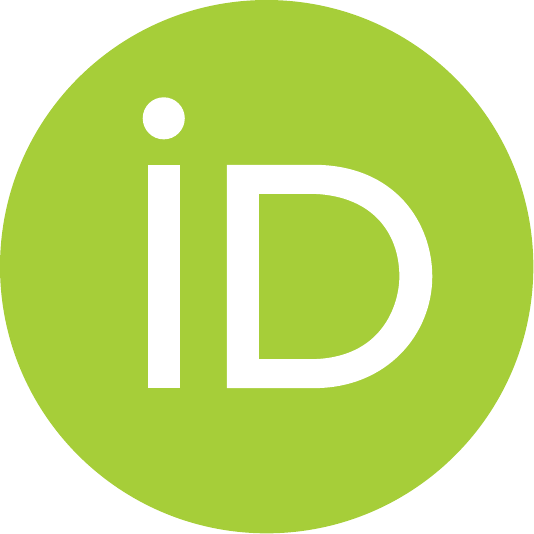}}}

\newcommand\K{\text{K}}

\begin{document}

\lefttitle{Publications of the Astronomical Society of Australia}
\righttitle{Kasiri et al.}

\jnlPage{1}{4}
\jnlDoiYr{2026}
\doival{10.1017/pasa.xxxx.xx}

\articletitt{Research Paper}

\title{Force convergence in Monte Carlo Lyman-$\alpha$ radiative transfer}

\author{Joshua Kasiri\orcids{0009-0009-6411-553X}$^{1}$, Aaron Smith\orcids{0000-0002-2838-9033}$^{1}$, Kevin Lorinc\orcids{0009-0005-3827-8774}$^{1}$, Olof Nebrin\orcids{0000-0003-3877-360X}$^{2}$, and Kazutaka Kimura\orcids{0000-0001-8382-3966}$^{3}$}

\affil{$^{1}$Department of Physics, The University of Texas at Dallas, Richardson, Texas 75080, USA}

\affil{$^{2}$Department of Astronomy \& Oskar Klein Centre for Cosmoparticle Physics, AlbaNova, Stockholm University, SE-106 91 Stockholm, Sweden}

\affil{$^{3}$Astronomical Institute, Graduate School of Science, Tohoku University, Aoba, Sendai 980-8578, Japan}

\corresp{J.\ Kasiri, Email: \href{mailto:Joshua.Kasiri@UTDallas.edu}{Joshua.Kasiri@UTDallas.edu}}

\begin{abstract}
  Monte Carlo radiative transfer (MCRT) is widely used to model Ly$\alpha$ resonant-line transport, but convergence is difficult to assess in optically thick media where photons undergo many scatterings before escape. This is especially important for internal quantities such as radiative acceleration and the force multiplier, which depend on momentum deposition throughout the gas rather than only on emergent spectra. We study the convergence of Ly$\alpha$ MCRT momentum-transfer estimators in static spherical clouds for both central point-source and uniform-source emission. We first establish diffusion-limit benchmarks for radial acceleration profiles and integrated force multipliers, then develop a moment-based framework for diagnosing convergence from the photon-packet contribution distribution. This framework separates three distinct questions: whether the estimator converges to the correct mean, how large its finite-sampling uncertainty is, and whether the estimated uncertainty is itself stable. We apply this hierarchy to the direct event-based scattering estimator, a gradient-of-energy-density estimator, and a divergence-of-radiation-pressure estimator. Zeroth-order convergence is assessed with profile comparisons, integrated force-multiplier bias, and finite-group relative error. First-order convergence is quantified with fractional error, the photon number required to reach a target precision, and the corresponding runtime requirement. Second-order convergence is tested with the coefficient of variation of variance, which measures the reliability of the variance estimate used in the first-order diagnostics. This organisation makes it clear that convergence of Ly$\alpha$ radiation forces is not captured by a single criterion. An estimator may be precise yet biased, unbiased yet computationally expensive, or apparently smooth while still having an unstable error estimate. Core-skipping prescriptions, source geometry, estimator construction, and spatial resolution enter this hierarchy in different ways. Our results provide a practical convergence framework for internal Ly$\alpha$ MCRT force calculations and show why statistical precision, computational cost, and physical accuracy must be evaluated separately.
\end{abstract}


\maketitle



\section{Introduction}
\label{sec:intro}

The emission and absorption of Lyman-$\alpha$ (Ly$\alpha$) photons provide one of the most important spectroscopic probes of the local and high-redshift Universe \citep{Hayes2015, Finkelstein2016, Ouchi2020}. Ly$\alpha$ is produced when ionizing radiation from young stars creates \HII regions and subsequent recombinations cascade through the $2p\rightarrow1s$ transition of neutral hydrogen. Because this transition is resonant, Ly$\alpha$ photons scatter multiple times before escape, encoding information about the gas distribution, kinematics, ionization state, and dust content of their source environments \citep{Partridge1967, Dijkstra2014}. The same resonant scattering process also transfers momentum to the gas. In low-dust environments, especially those expected around the first stars, galaxies, and massive black holes, this repeated momentum exchange can make Ly$\alpha$ radiation pressure a potentially dominant dynamical feedback channel \citep{Chandrasekhar1945, DijkstraLoeb2008, Smith2016, Smith2017, SmithDCBH2017, GeWise2017, Abe2018, Kimm2018, Tomaselli2021, Mushano2024, Ferrara2025, Manzoni2025, Nebrin2025, Nebrin2026, Menon2026}.

The predicted large forces and intensities from trapped Ly$\alpha$ photons could have important implications for the initial mass function of Population III stars \citep[e.g.][]{McKee2008, Jaura2022}, the formation of the first star clusters and their feedback-driven winds \citep{Abe2018, Kimm2018, Nebrin2025, Menon2026, Nebrin2026}, the low star formation efficiency observed in old Ultra-Faint Dwarf galaxies \citep{Nebrin2025}, and early black hole formation and growth \citep[e.g.][Kimura et al., in prep.]{GeWise2017, JohnsonDijkstra2017, Smith2016, Smith2017, Mushano2024}. Understanding Ly$\alpha$ transport is therefore crucial not only for interpreting observed spectra and surface-brightness profiles, but also for predicting internal radiation forces that may not be directly observable.

The difficulty is that Ly$\alpha$ radiative transfer is intrinsically non-local. Photons undergo both spatial and frequency diffusion, with the mean-free path changing rapidly as photons move between the core and wings of the line profile. Analytic solutions exist only for idealised geometries and limiting regimes, but they have provided much of the physical intuition for optically thick resonant-line transport \citep{Hummer1962, Osterbrock1962, Adams1972, Adams1975, Harrington1973, Neufeld1990}. Indeed, there has been renewed interest in the fundamental properties of Ly$\alpha$ radiative transfer reflected by sustained analytic progress \citep{LoebRybicki1999, Dijkstra2006, HansenOh2006, LaoSmith2020, Seon2020, Tomaselli2021, McClellan2022, Nebrin2025, Smith2025, Lorinc2025, LiZheng2026}. In the diffusion limit, the integrated momentum coupling is often summarised by the force multiplier $M_\mathrm{F}$, which measures the enhancement of the total force relative to the single-scattering value $\mathcal{L}/c$, for a source of Ly$\alpha$ luminosity $\mathcal{L}$ \citep[e.g.][]{DijkstraLoeb2008, Kimm2018, Smith2019, LaoSmith2020, Tomaselli2021, Nebrin2025, Nebrin2026}. For static spherical clouds, this quantity scales weakly with the line-centre optical depth as $M_\mathrm{F} \propto (a\tau_0)^{1/3}$ for both central point-source and uniform-source emission, with geometry-dependent normalisations \citep{LaoSmith2020}. These analytic benchmarks are invaluable, but realistic calculations generally require numerical radiative transfer.

Monte Carlo radiative transfer (MCRT) is a natural and highly accurate method for this problem because it follows the propagation of photon packets through a sequence of resonant scatterings without requiring the closure approximations used in direct treatments of the radiative transfer equation. Ly$\alpha$ MCRT codes with targeted optimisations have therefore been developed \citep{Ahn2002, Zheng2002, Dijkstra2006, Tasitsiomi2006b, Verhamme2006, Semelin2007, Laursen2009, Yajima2012, Smith2015, GronkeDijkstra2016, Seon2020, Michel-Dansac2020, Byrohl2025}, and applied in a wide range of settings, from idealised transfer problems to cosmological and galaxy-scale simulations \citep{Behrens2019, Laursen2019, Smith2019, Smith2022, Byrohl2021, Blaizot2023, Bhagwat2025}. More generally, Monte Carlo methods are widely used in radiative transfer because they are flexible, geometrically agnostic, and allow straightforward statistical estimators for both emergent observables and internal radiation-field quantities \citep{Lucy1999, Lucy2002, Wood2004, Kasen2006, Roth2015, NoebauerSim2019, Smith2020}. The main drawback is slow $\propto 1/\sqrt{N}$ convergence, so additional photon packets yield diminishing noise reduction. In optically thick media, photon histories can become long, heterogeneous, and expensive, so a large number of photon packets may be required before internal estimators are reliable \citep{Camps2018, Krieger2023, Krieger2024}.

For resonant-line radiative transfer, the standard way to reduce this cost is to use acceleration schemes such as core skipping. These methods explicitly avoid computing the many short mean-free-path scatterings that occur while a photon remains near line centre, instead pushing photons more efficiently into the wing where spatial transport becomes effective \citep{Ahn2002, Zheng2002, Dijkstra2006, Laursen2009, Smith2015, Michel-Dansac2020, Seon2020}. Such techniques are well motivated for emergent spectra and escape fractions, because core photons make little spatial progress before frequency diffusion carries them into the wing. However, core-skipping techniques compromise the accurate representation of the internal radiation field. The number of scatterings, trapping time, radiation energy density, and momentum deposition depend on what happens inside the cloud, so a computational acceleration method that preserves the escaping spectrum can still alter the internal radiation field. In a companion study, \citet{Lorinc2025}, we examined this issue from the spectral and physical side, identifying which frequency ranges contribute to force, trapping, and scattering statistics, and showing why missing core contributions cannot always be repaired by simple diffusion corrections.

The present paper addresses the statistical side of the same problem. Rather than asking only whether a simulated Ly$\alpha$ spectrum has converged, we ask how internal momentum-transfer estimators converge, how their sampling error should be measured, and how acceleration schemes change the value to which the simulation converges. This distinction is essential because different estimators can fail in qualitatively different ways. A direct scattering estimator measures the momentum exchanged at individual events, while path-based estimators infer force from radiation-field moments such as the energy density or radiation pressure. These estimators may have different bias, noise, spatial-resolution dependence, and computational cost, even when applied to the same photon ensemble.

We therefore develop a moment-based hierarchy for the convergence of Ly$\alpha$ MCRT force calculations. At zeroth order, we test whether an estimator converges to the correct mean profile or integrated force multiplier. At first order, we quantify finite-sampling noise through fractional error and the photon number required to reach a target precision. At second order, we ask whether the variance estimate itself is stable, using the coefficient of variation of variance. This framework separates physical accuracy from statistical precision and from the reliability of the quoted error bar.

The paper is organised as follows. In Section~\ref{sec:lyman-aplha_rt}, we summarise the Ly$\alpha$ transfer problem, define the point-source and uniform-source benchmark solutions, and validate the numerical calculations against newly derived diffusion-limit radial profiles. In Section~\ref{sec:convergence}, we introduce the statistical hierarchy of convergence and define the diagnostics used throughout the paper. In Section~\ref{sec:mcrt_convergence}, we apply these diagnostics to direct scattering, gradient-of-energy-density, and divergence-of-radiation-pressure force estimators, comparing no-core-skipping, fixed core-skipping, and dynamical core-skipping calculations. We summarise the implications for internal Ly$\alpha$ force calculations in Section~\ref{sec:conclusions}.

\begin{figure*}
    \centering
    \includegraphics[width=0.8\textwidth]{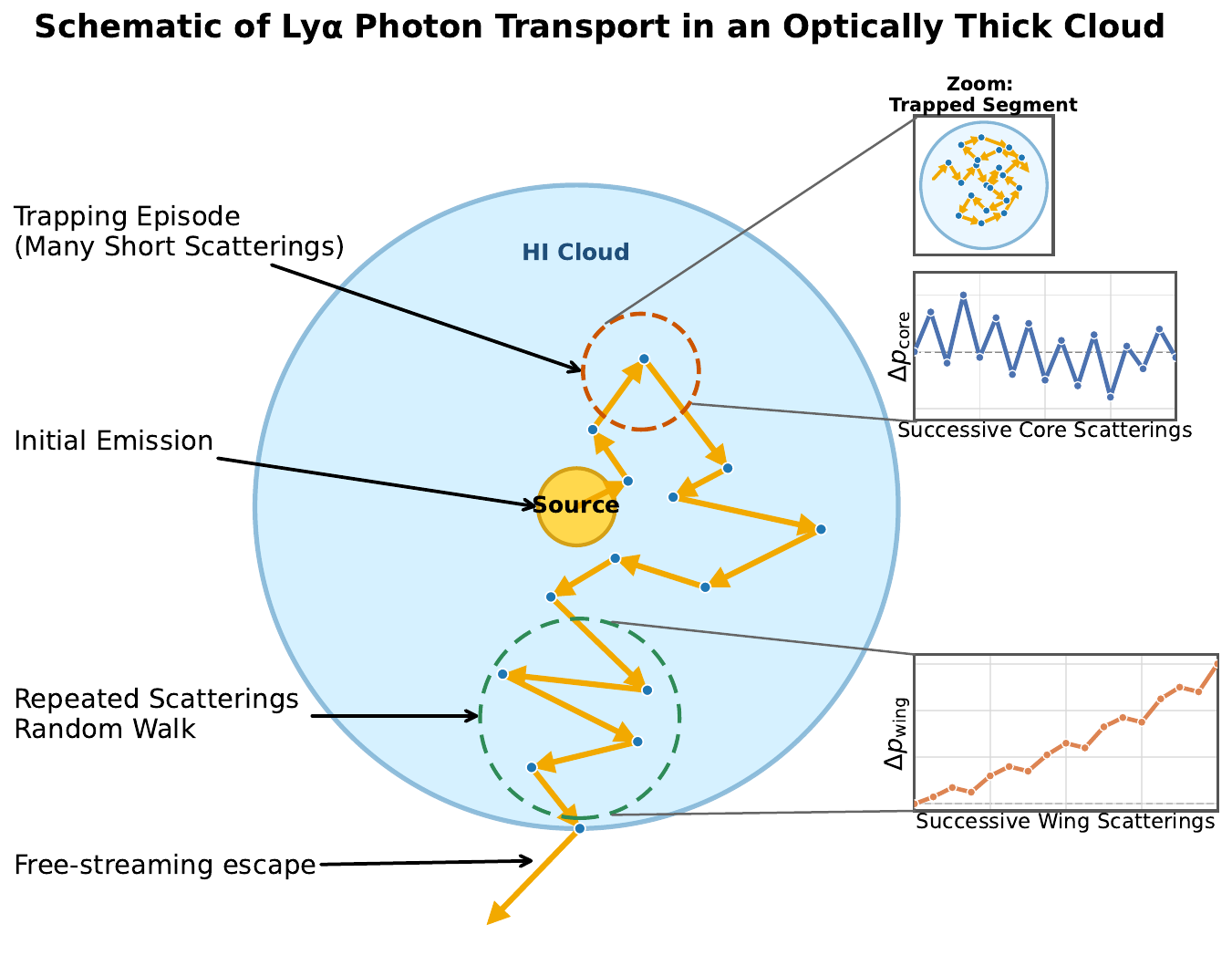}
    \caption{
    Schematic picture of Ly$\alpha$ photon transport in an optically thick, static neutral hydrogen cloud. Near line centre, photons undergo many short mean-free-path core scatterings, resulting in approximately isotropic momentum kicks with little net radial drift but significant nonzero variance. During rare wing excursions, the propagation distances are larger and photons can make coherent radially outward progress before either returning to the core or escaping the cloud. This motivates an initial heuristic picture in which core scatterings mainly contribute noise while wing excursions carry the mean momentum signal. However, the event-based acceleration estimator is a correlated sum over successive scattering events, so the variance is ultimately controlled by both the total number of kicks and their correlations along the photon trajectory.
    }
    \label{fig:lya_momentum_schematic}
\end{figure*}

\section{\texorpdfstring{Ly$\boldsymbol{\alpha}$ radiative transfer}{Lya radiative transfer}} \label{sec:lyman-aplha_rt}

In this section, we provide background on Ly$\alpha$ radiative transfer, focusing on aspects relevant for MCRT simulations. During transport, photons interact probabilistically with neutral hydrogen along their paths, with a mean-free path of $\lambda_\text{mfp} = 1/n_\HI \sigma_\nu$, where $n_\HI$ is the number density of neutral hydrogen and $\sigma_\nu$ is the frequency-dependent cross section. As photons undergo multiple scatterings, they randomly walk in both space and frequency, and the average path length between scatterings depends on frequency. To fully describe the steady-state radiation field, we need the specific intensity $I_\nu(\bm{r},\bm{n})$, which is the energy per unit frequency $\nu$ and propagation direction $\bm{n}$ as a function of spatial position $\bm{r}$. For a static medium with no dust, the steady-state radiative transfer equation is
\begin{equation} \label{eq:general_rt_equation}
  \bm{n} \bm{\cdot} \bm{\nabla} I_\nu = j_\nu - k_\nu I_\nu + \iint k_{\nu'} I_{\nu'} R_{\nu', \bm{n}' \rightarrow \nu, \bm{n}} \, \text{d}\Omega' \text{d}\nu' \, ,
\end{equation}
where $j_\nu$ and $k_\nu$ are the emission and absorption coefficients, respectively. The integral in the last term on the right accounts for frequency redistribution caused by partially coherent scattering \citep{Dijkstra2014}. The redistribution function $R_{\nu', \bm{n}' \rightarrow \nu, \bm{n}}$ is the differential probability per unit initial photon frequency $\nu'$ and per unit initial directional solid angle $\Omega'$ that scattering of such a photon travelling in the direction $\bm{n}'$ would place the scattered photon at frequency $\nu$ and in the direction of the unit vector $\bm{n}$. For the remainder of the paper, we convert to dimensionless frequency
\begin{equation} \label{eq:x}
  x \equiv \frac{\nu - \nu_0}{\Delta \nu_\text{D}} \, ,
\end{equation}
where $\nu_0 = 2.466 \times 10^{15} \, \text{Hz}$ is the frequency at line centre for Ly$\alpha$, $\Delta \nu_\text{D} \equiv (v_\text{th}/c)\nu_0$ the Doppler width of the profile, and $v_\text{th} \equiv (2 k_\text{B} T / m_\text{H})^{1/2}$ the thermal velocity. To characterise the frequency dependence on the cross section, we define the Hjerting--Voigt function $H(a,x) = \sqrt{\pi} \Delta \nu_\text{D} \phi_\text{Voigt}(\nu)$ as the dimensionless convolution of Lorentzian and Maxwellian distributions,
\begin{equation} \label{eq:H}
  H(a,x) = \frac{a}{\pi} \int_{-\infty}^\infty \frac{e^{-y^2}\text{d}y}{a^2+(y-x)^2} \approx
    \begin{cases}
      e^{-x^2} & \quad \text{`core'} \\
      {\displaystyle \frac{a}{\sqrt{\pi} x^2} } & \quad \text{`wing'}
    \end{cases} \, .
\end{equation}
The `damping parameter', $a \equiv \Delta \nu_{\rm L} /2 \Delta \nu_{\rm D}$, describes the relative broadening compared to the natural line width, which is $\Delta \nu_\text{L} = 9.936 \times 10^7 \, \text{Hz}$ for Ly$\alpha$. Focusing on Ly$\alpha$ from now on, in an isothermal gas, $a = 4.702 \times 10^{-4} \, (T/10^4 \, \K)^{-1/2}$. The cross section at dimensionless frequency $x$ is $\sigma_x = \sigma_0 H(a,x)$, where $\sigma_0 = 5.898 \times 10^{-14} \, (T/10^4 \, \K)^{-1/2} \, \rm cm^2$ is the cross section at line centre. This allows us to give the optical depth as $\tau_x = \int_\text{path} k(\ell) \, H(a,x) \, \text{d}\ell$, where $k(\bm{r}) = n_\HI(\bm{r}) \sigma_0$ is the absorption coefficient at line centre at position $\bm{r} = \bm{r}_0 + \ell \bm{n}$.

It is possible to solve Eq.~(\ref{eq:general_rt_equation}) for the angular-averaged frequency $J_x \equiv \frac{1}{4\pi} \int I_x \, \text{d} \Omega$, where $J_x = \Delta \nu_\text{D} \, J_\nu$, in the diffusion approximation, which is valid when $a \tau_0 \gg 10^3$ \citep[for a comprehensive discussion see][]{Nebrin2025}. Then one can derive the radiation energy density, related to the angular-averaged intensity as $u(\bm{r}) = \frac{4\pi}{c} \int J_x(\bm{r})\,\text{d}x$, the trapping time, defined as $t_\text{trap} \equiv \mathcal{L}^{-1} \int u(\bm{r})\,\text{d}V$ with $\mathcal{L}$ denoting the source luminosity, and the outward force multiplier
\begin{equation} \label{eq:sec2_M_F}
  M_\text{F}
  \equiv \mathcal{L}^{-1}  \iint k(\bm{r}) F \,\text{d}x\,\text{d}V
  \approx -\frac{c}{3\mathcal{L}}  \int \nabla u(\bm{r})\,\text{d}V \, ,
\end{equation}
where the second expression follows from Fick's Law replacing the flux by $F \approx -c \nabla u / 3k$ for diffusive radiation. The force multiplier quantifies the enhancement of the momentum coupling compared to the single scattering limit of $\mathcal{L}/c$. This is related to the radial force per unit volume relative to $\mathcal{L}/c$ by:
\begin{equation}
  M_\text{F} = \int \frac{\rho a(r)}{\mathcal{L}/c} \, \text{d}V \approx (a \tau_0)^{1/3} \times \begin{cases} 3.51 & \text{Central point source} \\ 0.51 & \text{Uniform source} \end{cases} ,
\end{equation}
where the last expression is the solution for the bracketing cases of central point and uniform sources within a static, homogeneous spherical cloud \citep{LaoSmith2020}.\footnote{The smaller constant pre-factor for the uniform source case ($\approx 0.51$) is due to flux cancellation, which lowers the net radial force.} We now generalise these results for radial force \textit{profiles}, which to our knowledge have not appeared in previous studies. For brevity, we omit the full derivations but the closed-form expressions follow by substituting the appropriate series for the radiation field $J_x(\bm{r})$ from \cite{LaoSmith2020} into Eq.~(\ref{eq:sec2_M_F}). All expressions are real-valued upon evaluation and are validated numerically below.

\subsection{Central point source}
The radial acceleration for a point source is
\begin{align} \label{eq:a_point(r)}
  \frac{\rho a(\tilde{r})}{\mathcal{L}/c} &= -\frac{\Gamma(4/3)}{2 \pi^{4/3} R^3} \left(\frac{2 a \tau_0}{\sqrt{\pi}}\right)^{1/3} \notag \\
  &\times \left( \pi \frac{\text{Re}[\text{Li}_{-2/3}(e^{i \pi \tilde{r}})]}{\tilde{r}} - \frac{\text{Im}[\text{Li}_{1/3}(e^{i \pi \tilde{r}})]}{\tilde{r}^2} \right) \, ,
\end{align}
where $\text{Li}_s(z)$ is the polylogarithm function $\text{Li}_s(z)=\sum_{n=1}^{\infty}z^n/n^s$. The cumulative contribution within a radius $\tilde{r} \equiv r / R$ is
\begin{align} \label{eq:M_F_r}
  &\frac{M_\text{F}(<r)}{M_\text{F}} \equiv \frac{\int_0^r a(r) \, r^2 \, \text{d}r}{\int_0^R a(r) \, r^2 \, \text{d}r} \notag \\
  &\qquad = \frac{2\,\text{Re}[\text{Li}_{4/3}(e^{i \pi \tilde{r}})] + \pi\,\tilde{r}\,\text{Im}[\text{Li}_{1/3}(e^{i \pi \tilde{r}})] - 2 \zeta(4/3)}{(2^{2/3} - 4) \zeta(4/3)} \, .
\end{align}
For completeness, we also provide equivalent formulae for the energy density \citep[see eq.~89 from][]{LaoSmith2020}:
\begin{equation}  \label{eq:u_point(r)}
  u(\tilde{r}) = \frac{\mathcal{L}}{c R^2} \Gamma\left(\frac{1}{3}\right) \frac{(2 a \tau_0)^{1/3}}{2 \pi^{3/2}} \tilde{r}^{-1} \text{Im}\left[\text{Li}_{1/3}\left(e^{i \pi \tilde{r}}\right)\right] \, ,
\end{equation}
which has a cumulative radial contribution of
\begin{align} \label{eq:t_trap_r}
  \frac{t_\text{trap}(<r)}{t_\text{trap}} &\equiv \frac{\int_0^r u(r) \, r^2 \, \text{d}r}{\int_0^R u(r) \, r^2 \, \text{d}r} \notag \\
  &= \frac{\pi\,\tilde{r}\,\text{Re}[\text{Li}_{4/3}(e^{i \pi \tilde{r}})] - \text{Im}[\text{Li}_{7/3}(e^{i \pi \tilde{r}})]}{(2^{-1/3} - 1) \pi\,\zeta(4/3)} \, .
\end{align}
The number density of scatterings is
\begin{equation}
  \rho_\text{scat}(\tilde{r}) = \frac{\tau_0 \sqrt{6}}{8 \sqrt{\pi} R^3} \frac{\cot(\pi \tilde{r}/2)}{\tilde{r}} \, ,
\end{equation}
which has a cumulative radial contribution of
\begin{align} \label{eq:N_scat_r}
  &\frac{N_\text{scat}(<r)}{N_\text{scat}} \equiv \frac{\int_0^r \rho_\text{scat}(r) \, r^2 \, \text{d}r}{\int_0^R \rho_\text{scat}(r) \, r^2 \, \text{d}r} \notag \\
  &\qquad = \frac{\pi (3 \tilde{r}^2 - 2) + 12 i \tilde{r} \log(1 - e^{i \pi \tilde{r}}) +
    12\,\text{Li}_2(e^{i \pi \tilde{r}})/\pi}{6 i \log(4)} \, .
\end{align}

\begin{figure*}
    \centering
    \includegraphics[width=\textwidth]{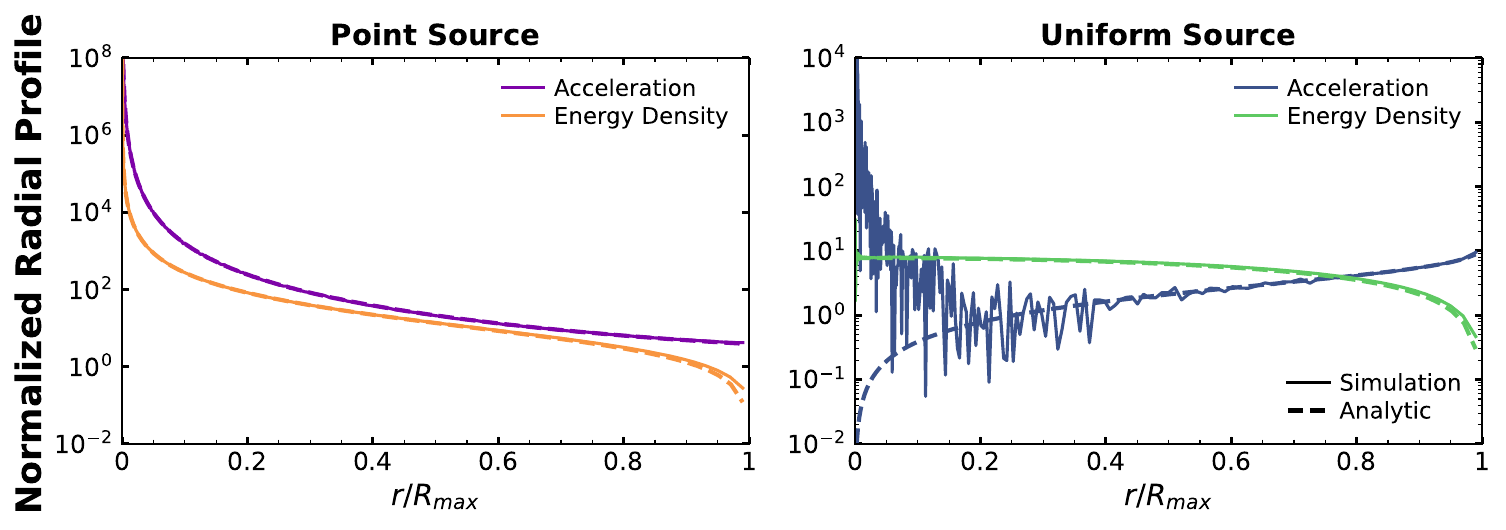}
    \caption{Numerical validation of the analytic diffusion-limit solutions for a static spherical cloud with $\tau_0=10^8$, $x_\mathrm{crit} = 0$, and $N_\mathrm{ph} = 10^6$. \textit{Left:} Point-source radial profiles, normalised as $R^3\rho a(r)/(\mathcal{L}/c)$ and $c R^2 u(r) / \mathcal{L}$ for acceleration and energy density respectively, as given by Eqs.~(\ref{eq:a_point(r)}) and (\ref{eq:u_point(r)}). \textit{Right:} Uniform-source radial profiles, normalised the same as point-source, as given by Eqs.~(\ref{eq:a_uniform(r)}) and (\ref{eq:u_uniform(r)}). The dashed curves show the analytic solutions and solid curves show the MCRT results for central point-source and uniform-source emission. At sufficiently large optical depth, the simulations reproduce the analytic profiles for both quantities and both source geometries. The only major visible deviation occurs for the uniform-source acceleration near the geometric centre, where the estimator becomes noisy because few photons are launched in the small-volume central region.}
    \label{fig:validation}
\end{figure*}

\subsection{Uniform source}
The radial acceleration for a uniform source is
\begin{align} \label{eq:a_uniform(r)}
  \frac{\rho a(\tilde{r})}{\mathcal{L}/c} &= \frac{3\Gamma(4/3)}{2 \pi^{10/3} R^3} \left(\frac{2 a \tau_0}{\sqrt{\pi}}\right)^{1/3} \notag \\
  &\times \left( \pi \frac{\text{Re}[\text{Li}_{4/3}(-e^{i \pi \tilde{r}})]}{\tilde{r}} - \frac{\text{Im}[\text{Li}_{7/3}(-e^{i \pi \tilde{r}})]}{\tilde{r}^2} \right) \, .
\end{align}
The cumulative contribution within a radius $\tilde{r} \equiv r / R$ is
\begin{align} \label{eq:M_F_r_uniform}
  &\frac{M_\text{F}(<r)}{M_\text{F}} \equiv \frac{\int_0^r a(r) \, r^2 \, \text{d}r}{\int_0^R a(r) \, r^2 \, \text{d}r} \notag \\
  &= \frac{1}{(16 - 2^{2/3}) \zeta(10/3)} \bigg(8\,\text{Re}[\text{Li}_{10/3}(-e^{i \pi \tilde{r}})] \notag \\
  &\quad\qquad + \,4\pi\,\tilde{r}\,\text{Im}[\text{Li}_{7/3}(-e^{i \pi \tilde{r}})]+(8 - 2^{2/3}) \zeta(10/3)\bigg)  \, .
\end{align}
The energy density is \citep[see eq.~98 from][]{LaoSmith2020}
\begin{equation} \label{eq:u_uniform(r)}
  u(\tilde{r}) = -\frac{3\mathcal{L}}{c R^2} \Gamma\left(\frac{1}{3}\right) \frac{(2 a \tau_0)^{1/3}}{2 \pi^{7/2}} \tilde{r}^{-1} \text{Im}\left[\text{Li}_{7/3}\left(-e^{i \pi \tilde{r}}\right)\right] \, ,
\end{equation}
which has a cumulative radial contribution of
\begin{align} \label{eq:t_trap_r_uniform}
  \frac{t_\text{trap}(<r)}{t_\text{trap}} &\equiv \frac{\int_0^r u(r) \, r^2 \, \text{d}r}{\int_0^R u(r) \, r^2 \, \text{d}r} \notag \\
  &= \frac{\pi\,\tilde{r}\,\text{Re}[\text{Li}_{10/3}(-e^{i \pi \tilde{r}})] - \text{Im}[\text{Li}_{13/3}(-e^{i \pi \tilde{r}})]}{\pi\,\zeta(10/3)} \, .
\end{align}
The number density of scatterings is
\begin{equation}
  \rho_\text{scat}(\tilde{r}) = \frac{3 \tau_0 \sqrt{6}}{4 \pi^{5/2} R^3} \tilde{r}^{-1} \text{Im}\left[\text{Li}_2\left(-e^{-i \pi \tilde{r}}\right)\right] \, ,
\end{equation}
which has a cumulative radial contribution of
\begin{align} \label{eq:N_scat_r_uniform}
  &\frac{N_\text{scat}(<r)}{N_\text{scat}} \equiv \frac{\int_0^r \rho_\text{scat}(r) \, r^2 \, \text{d}r}{\int_0^R \rho_\text{scat}(r) \, r^2 \, \text{d}r} \notag \\
  &\qquad = \frac{\pi\,\tilde{r}\,\text{Re}[\text{Li}_{3}(-e^{i \pi \tilde{r}})] - \text{Im}[\text{Li}_{4}(-e^{i \pi \tilde{r}})]}{\pi\,\zeta(3)} \, .
\end{align}

\subsection{Numerical validation}
To solve Eq.~(\ref{eq:general_rt_equation}) directly, we employ the MCRT algorithm as implemented in the Cosmic Ly$\alpha$ transfer code \citep[\textsc{colt};][]{Smith2015, Smith2019, SmithMW2022, McClymont2025}, which samples individual photon paths to generate statistically converged properties of the radiation field. In brief, the optical depth traversed between scattering events is drawn from an exponential distribution. Once the photon has travelled far enough to attenuate this randomly drawn optical depth, it undergoes a scattering event, and its frequency and direction are updated accordingly. This process is repeated until the photon escapes the computational domain. With enough photon paths, the method yields converged radiation-field properties, solving Eq.~(\ref{eq:general_rt_equation}) for the intensity $I(\bm{r},x,\bm{n})$ and related quantities with the definitions provided.

Although accurate when fully converged without using any acceleration scheme, resonant-line MCRT suffers significantly from repeated core scatterings when photons are near line centre. It is standard practice to use core-skipping algorithms to preferentially interact with atoms whose velocities are large enough to push photons into the wings of the line profile. Core-skipping retains accuracy for emergent properties, because photons make little spatial progress until frequency diffusion places them in the wing. However, core photons can affect internal quantities, especially the radiative force, because repeated core scatterings impart momentum in a way that does not completely cancel out. We quantify the losses from core-skipping below. We also explore the detailed convergence properties of Ly$\alpha$ MCRT, which depend on the number of scattering events. This number scales with the line-centre optical depth as $\sim\tau_0$ without core-skipping or as $\tau_\text{esc}^2 \sim (a\tau_0)^{2/3}$ with core-skipping \citep{Smith2018, Lorinc2025}.

As a final validation of these benchmarks, we compare the MCRT profiles directly with the analytic solutions derived above in Fig.~\ref{fig:validation}. At sufficiently large optical depth, where the diffusion-limit solutions are applicable, the simulations closely reproduce the analytic profiles. The agreement is essentially exact for the energy density and for the point-source acceleration profile. The only visible exception is the uniform-source acceleration near the geometric centre of the cloud, where the simulated profile becomes noisy. This is a sampling limitation of MCRT rather than a failure of the analytic solution. For unbiased uniform emission, very few photons are launched in the small-volume central region. This issue is discussed in more detail in Appendix~\ref{app:uniform_volume_boosting}.

\subsection{Scaling expectations for momentum fluctuations}
\label{sec:momentum_scaling}

Before introducing convergence diagnostics, it is useful to estimate how the variance of the event-based momentum-kick estimator should scale with optical depth. This is only intended as a scaling argument, not a closed-form statistical theory. The mean force multiplier is already fixed by the diffusion-limit solution,
\begin{equation}
    \langle Y\rangle \sim M_\text{F} \sim (a\tau_0)^{1/3} \, ,
\end{equation}
where $Y$ denotes the dimensionless contribution per-photon to the total radial momentum deposition, normalised by the single-scattering momentum, $p_\gamma = h\nu_0/c$.

For an individual scattering event, denoted by $e$, the radial momentum deposited into the gas is approximately
\begin{equation}
    \Delta p_e \simeq p_\gamma (\mu_{e} - \mu_{e-1}) \, ,
\end{equation}
where $\mu \equiv \bm{n} \cdot \hat{\bm{r}}$ and $\mu_e$ represent the outgoing radial photon direction after the event, while $\mu_{e-1}$ represents the incoming radial photon direction prior to the event. In the core, the photon direction can be taken to be nearly isotropic, so $\langle \mu\rangle \simeq 0$ and $\langle \mu^2\rangle \simeq 1/3$. Thus, core scatterings contribute little to the mean radial force in this approximation, but each scattering still contributes an order-unity kick variance. If all kicks are independent, then
\begin{equation} \label{eq:sigma_Y_Nscat}
    \sigma_Y^2 \sim N_\text{scat} \, ,
\end{equation}
and since the number of scatterings in an extremely optically thick static medium scales approximately as $N_\text{scat}\sim \tau_0$ \citep{Osterbrock1962,Adams1972,Adams1975}, one would obtain
\begin{equation}
    \sigma_Y \sim \tau_0^{1/2} \, .
\end{equation}
Compared with the mean, $\langle Y\rangle \sim (a\tau_0)^{1/3}$, this gives the independent-kick expectation for the coefficient of variation of
\begin{equation}
    \mathrm{CV} \equiv \frac{\sigma_Y}{\langle Y\rangle}
    \sim a^{-1/3}\tau_0^{1/6} \, .
\end{equation}
This heuristic picture is summarised schematically in Fig.~\ref{fig:lya_momentum_schematic}. Core scatterings are frequent and nearly isotropic, so they contribute little to the mean radial momentum deposition in this simple estimate, while still producing appreciable fluctuations in the accumulated momentum between photon realisations. By contrast, wing excursions have longer mean free paths and are responsible for the escape-scale transport that sets the mean force multiplier. In this initial picture, the wing therefore supplies the signal, while the core supplies much of the noise.

However, the event-based force estimator is \textit{not} a sum of independent random kicks. The photon trajectory is correlated, with different kicks and frequencies depending on each other. Covariance can therefore separate the raw variance generated by individual kicks from the final variance of the estimator. For a sequence of dimensionless kicks, $\varrho_e = \mu_e - \mu_{e-1}$, the variance of a single photon history is
\begin{equation}
    \mathrm{Var}(Y)
    =
    \sum_e \mathrm{Var}(\varrho_{e})
    +
    2\sum_{e<f}\mathrm{Cov}(\varrho_{e},\varrho_{f}) \, .
\end{equation}
This covariance sum leads to consequential differences between a true event-based momentum estimator and an independent-kick random walk. Exploring this covariance in depth is beyond the scope of this paper. Instead, we will analyse photon-level covariance in a companion paper (Kasiri et al., in prep). Let us take the raw variance stemming from the direct kicks to be
\begin{equation}
    \sigma_\mathrm{kick}^2 = \sum_e \mathrm{Var}\left( \varrho_{p,e} \right) \, ,
\end{equation}
and the level of reinforcement or suppression from the covariance as $\mathrm{R}$. The final variance of the estimator should then scale as
\begin{equation}
    \sigma_Y^2 \sim \mathrm{R} \sigma_\mathrm{kick}^2 \, .
\end{equation}
The independent sequence scaling form of Eq.~(\ref{eq:sigma_Y_Nscat}) naturally applies to the direct kick variance, $\sigma_\mathrm{kick}^2 \sim N_\mathrm{scat}$. What matters is then understanding how this variance is suppressed or reinforced.

The covariance term is expected to act primarily as a suppression of variance. This is due to the telescoping nature of the momentum kicks, since the incoming momentum state for one event is the outgoing momentum state of the prior event. This can be directly shown by adding two dimensionless kicks,
\begin{equation}
    \varrho_e + \varrho_{e+1} = (\mu_{e} - \mu_{e-1}) + (\mu_{e+1} - \mu_e) = \mu_{e+1} - \mu_{e-1} \, .
\end{equation}
In the idealised limit, all intermediate momentum terms telescope away, leaving the deposited momentum controlled by boundary terms. In the actual spherical transport problem, this cancellation is incomplete because the photon's position, frequency, and direction evolve with each scattering event. The intermediate terms do not vanish completely. Instead, they are diminished relative to the independent-kick picture. This implies that the large raw kick variance generated by many core scatterings is greatly reduced through negative covariance. As a result, the total variance of the estimator is set by the transport scale associated with escape rather than by the raw number of scatterings.

It is important to distinguish this intra-history covariance from inter-packet covariance. The covariance terms above refer to correlations between scattering events within a single photon trajectory. These correlations determine the variance of the per-packet momentum contribution, $Y$, and therefore may change the optical-depth scaling of $\sigma_Y^2$. In contrast, different photon packets are launched and propagated independently in the MCRT calculation. Thus, once each photon history has been compressed into a single contribution $Y_p$, distinct photon packets satisfy
\begin{equation}
    \mathrm{Cov}(Y_p,Y_q) \approx \mathrm{Var}(Y_p) \, \delta_{pq} \, ,
\end{equation}
using the Kronecker delta function $\delta_{pq}$. The scaling argument below therefore allows for correlated scatterings within each photon history, while still treating different photon histories as independent Monte Carlo samples.

The natural Ly$\alpha$ scale for the total variance is set by the escape frequency. In the wing, $H(a,x)\propto a/x^2$, and the usual excursion argument gives
\begin{equation}
    x_\text{esc}\sim (a\tau_0)^{1/3} \, ,
\end{equation}
the frequency at which a photon can make an escape-scale spatial excursion before returning to line centre \citep{Adams1975, Harrington1973, Neufeld1990, Dijkstra2014}. We therefore introduce an ansatz for the parametrisation of the final variance of the momentum-kick estimator as
\begin{equation}
    \sigma_Y^2 \sim \left( a \tau_0 \right)^{2/3} \, .
\end{equation}
Since $\langle Y\rangle\sim(a\tau_0)^{1/3}$, the variance and signal are both set by the transport scale. So the corresponding coefficient of variation is
\begin{equation}
    \mathrm{CV}
    \equiv
    \frac{\sigma_Y}{\langle Y\rangle}
    \sim
    \mathrm{constant} \, .
\end{equation}
This means that the per-photon contribution distribution does not depend on optical depth. If we average over $N_\mathrm{photons}$ independent photon histories, then we get a fractional error of
\begin{equation}
    \mathrm{FE}
    \sim
    \frac{\mathrm{CV}}{\sqrt{N_\mathrm{ph}}} \sim N_\mathrm{ph}^{-1/2} \, ,
\end{equation}
and therefore the photon number required to reach a target fractional error scales as
\begin{equation}
    N_\text{needed}
    \sim \frac{\mathrm{CV}^2}{\mathrm{FE}_\mathrm{target}^2} \sim \mathrm{FE}_\mathrm{target}^{-2} \, .
\end{equation}

This scaling predicts no clear optical-depth dependence for the convergence metrics. That may at first seem counterintuitive. Changing the optical depth changes how strongly the medium interacts with the radiation, which should affect the fluctuations. However, the suppression of fluctuations may also grow with optical depth. The more opaque the medium, the closer the physical setup comes to the ideal case where the photon's position, frequency, and direction are constant and momentum kicks telescope completely. If both the fluctuations and their suppression scale in the same manner, then no clear optical-depth dependence will emerge.

It is important to note that these scalings do $\it{not}$ imply universal values for FE, CV, or $N_\mathrm{needed}$. These quantities can still vary across geometries, source types, and algorithms. The scaling argument only states that the opacity of the system should not set them directly. These scalings provide a useful interpretive framework for the convergence results below. In particular, the observed dependence of FE and $N_\text{needed}$ on $\tau_0$ can be read as an empirical probe of how accurate the picture of telescoping momentum kicks really is.

\section{Hierarchy of convergence} \label{sec:convergence}

We now develop a hierarchy for analysing the statistical convergence of Ly$\alpha$ MCRT simulations. Each sampled photon provides a noisy, partial estimate of the internal radiation field and associated forces. The resulting profiles or summary statistics are often treated as the sole ``answer'' for a given simulation. However, the underlying processes in MCRT are inherently stochastic, so any single realisation represents a random draw from a broader distribution of outcomes. To regard a result as physically reliable in this case, we need to understand the mean behaviour, the statistical fluctuations, and the stability of the error estimate itself.

The central difficulty is that the true underlying radiation field is unknown, because that would require an exact solution of Eq.~(\ref{eq:general_rt_equation}). As mentioned previously, approximate solutions can only be derived for certain idealised configurations, such as uniform clouds in the diffusion limit of large $\tau_0$. In the general case of arbitrary 3D geometries, numerical MCRT simulations provide the most reliable predictions. However, it is not always feasible to realise brute-force convergence without acceleration schemes, and these can introduce biases that degrade the accuracy to a level comparable to unconverged results. Without proper benchmark comparisons, self-convergence must be analysed statistically, using only the photon contributions generated within a simulation.

A naive approach to convergence would be to increase the number of photons until target metrics stabilise and successive runs appear similar to one another. This procedure is appropriate for deterministic methods, but Ly$\alpha$ MCRT photon histories can be highly heterogeneous, making it less robust when results are sensitive to outliers. A small subset of long-lived photons may dominate the signal, and the resulting distributions of contributions can be broad or heavy-tailed, with large differences between mean, median, or weighted statistics. The raw photon count $N$ can therefore be a poor proxy for statistical quality, and the smoothness of a given profile can conceal underlying residual noise or systematic bias.

To address this, we adopt a moment-based framework that treats the collection of photon-packet contributions as a statistical ensemble. Each photon packet contributes a completed value $Y_p$ to a chosen estimator, such as the contribution to radiative acceleration in a given radial shell or to a volume-integrated force multiplier. The distribution of these $Y_p$ values is described by its statistical moments. The first moment captures the total or mean signal, the second moment captures the power in fluctuations, and the higher moments quantify the stability of those fluctuations under resampling. Expressing diagnostics in terms of these moments provides an internally consistent way to measure convergence across different physical quantities and estimator choices.

Within this framework, we organise convergence into a hierarchy of orders. At zeroth order, we focus on the spatial profiles and global values that represent the physical quantities themselves. At this level, we also consider how the estimated mean approaches its asymptotic or expected values as we increase the number of photons, allowing us to characterise potential bias. At first order, we quantify the variance of the estimator, or how much the Monte Carlo result is expected to fluctuate due to finite sampling. This naturally leads to metrics such as fractional error, which connects the first and second moments to the precision of a simulation. At second order, we examine the coefficient of variation of variance. This reveals how stable our error estimates are when data are resampled. This second-order perspective becomes especially crucial when contributions are highly uneven or heavy-tailed, as is often the case for Ly$\alpha$ radiative transfer.

In the following subsections, we first formalise the statistical language needed for this analysis. We introduce the relevant moments, Monte Carlo estimators, and convergence metrics. We then build up the hierarchy of convergence diagnostics, showing how quantities such as relative error, fractional error, and the coefficient of variation, extended to variance as well, arise from the same underlying moment structure and can be applied to assess the reliability of Ly$\alpha$ MCRT simulation results.

\subsection{Statistical moments and Monte Carlo estimators}

The foundation of our convergence analysis is the interpretation of photon contributions as samples drawn from an underlying statistical distribution. Every MCRT simulation produces a large number of photon histories, and each history contributes some amount to the estimator of interest. These contributions differ from photon to photon because each random walk samples a different sequence of optical depths, frequency redistributions, and angular phase-function draws. For simplicity, we do not consider the impact of dust and other destruction mechanisms in this work.

Formally, for a chosen estimator, the completed packet contributions $\{Y_1,Y_2,\ldots,Y_{N_\mathrm{ph}}\}$ form a finite sample drawn from an unknown contribution distribution $P(Y)$, determined by the physics of radiative transfer and the simulation setup. We do not know the exact statistical moments of $P(Y)$, so we estimate them from the photon samples themselves. The resulting Monte Carlo estimators therefore inherit statistical noise that must be characterised.

\subsubsection{Raw moments (photon sums)}

It is natural to begin with raw moments since these are the quantities most closely related to what the code tallies. For an event-based estimator, the contribution from a single photon packet may itself be a sum over many events along that photon history. We therefore write the completed contribution of the photon packet $p$ as
\begin{equation}
    Y_p = C \sum_{e \in p} y_{p,e} \, ,
\end{equation}
where $y_{p,e}$ is the elementary event contribution and $C$ is the physical normalisation required for the estimator of interest. For example, for the scattering-based radial acceleration estimator, $C=(\mathcal{L}/c)/\rho$, while the individual event contributions contain the photon weight and the radial momentum factor, $y_{p,e}\sim w_{p,e}(\mu_{p,e}-\mu_{p,e-1})$. In Section~\ref{sec:momentum_scaling}, $Y$ denotes the underlying per-photon contribution random variable, while $Y_p$ denotes the realised contribution from the photon packet $p$ here.

The statistical sample used for convergence is then the set of completed photon-packet contributions $\{Y_1,Y_2,\ldots,Y_{N_\mathrm{ph}}\}$. The raw moment of order $k$ is
\begin{equation}
    S_k \equiv \sum_{p=1}^{N_\mathrm{ph}} Y_p^k \, .
\end{equation}
In spatially resolved calculations, the same definition applies shell by shell, with $Y_{p,i}$ denoting the contribution of packet $p$ to shell $i$. This notation is useful because it separates the two levels of randomness. The many scattering events within a single photon history may be correlated, and those correlations determine the distribution of $Y_p$. However, distinct photon packets are independent Monte Carlo samples of that distribution.

The first raw moment,
\begin{equation}
    S_1 = \sum_{p=1}^{N_\mathrm{ph}} Y_p \, ,
\end{equation}
is the estimator for the signal. The second raw moment,
\begin{equation}
    S_2 = \sum_{p=1}^{N_\mathrm{ph}} Y_p^2 \, ,
\end{equation}
measures the power in the packet-to-packet fluctuations. Since the estimators used here are linear sums of photon contributions, the raw moments provide the most direct statistical objects for analysing precision and convergence.

\subsubsection{Mean moments (photon averages)}

For comparison with standard statistical quantities, we also define per-photon sample moments,
\begin{equation}
    \mu_k \equiv \frac{S_k}{N_\mathrm{ph}} \, .
\end{equation}
The corresponding sample estimate of the packet-level variance is
\begin{equation}
    \sigma_Y^2 = \mu_2-\mu_1^2 \, .
\end{equation}
These quantities are useful when connecting the MCRT diagnostics to classical statistics such as the coefficient of variation.

\subsubsection{Variance of a MC estimator}

The variance of the first moment is simply \begin{equation} \mathrm{Var} \left( S_1 \right) = \sum_{p=1}^{N_\mathrm{ph}} \mathrm{Var} \left( Y_p \right) = N_\mathrm{ph} \sigma_Y^2 \, \end{equation} since the photon packets are drawn independently. Using the sample moments just defined gives \begin{equation} \mathrm{Var} \left( S_1 \right) = N_\mathrm{ph} \left( \frac{S_2}{N_\mathrm{ph}} - \frac{S_1^2}{N_\mathrm{ph}^2} \right) = S_2 - \frac{S_1^2}{N_\mathrm{ph}} \, .
\end{equation}

Although our convergence diagnostics are expressed in terms of completed photon-packet contributions, $Y_p$, they do not assume statistical independence among the individual events within a photon history. Covariance within each history is instead accounted for by decomposing the completed packet contribution into partial contributions from different regions of the trajectory. For example, we may write
\begin{equation}
Y_{p} = \sum_{\alpha} Y^{(\alpha)}_{p} \, ,
\end{equation}
where $\alpha$ labels some partition of the packet history, such as frequency range, spatial region, or another physically meaningful subdivision. We then define the corresponding partial first moments,
\begin{equation}
S_{1}^{(\alpha)} = \sum_{p=1}^{N_\mathrm{ph}} Y_{p}^{(\alpha)} \, ,
\end{equation}
and the raw cross-moment matrix,
\begin{equation}
M_{\alpha\beta} = \sum_{p=1}^{N_{\mathrm ph}} Y_{p}^{(\alpha)} Y_{p}^{(\beta)} \, .
\end{equation}
Subtracting the product of the sample means gives the covariance contribution matrix,
\begin{equation}
C_{\alpha\beta} = M_{\alpha\beta} - \frac{S_1^{(\alpha)} S_1^{(\beta)}}{N_{\mathrm ph}} \, .
\end{equation}
To obtain the final variance of the estimator, we then sum over all partial history indices
\begin{equation}
\sum_{\alpha,\beta} C_{\alpha\beta} = S_{2} - \frac{S_{1}^{2}}{N_{\mathrm ph}} =
\mathrm{Var} \left( S_1 \right) \, .
\end{equation}

Thus, the variance used in the convergence metrics automatically includes both the variance of the individual partial contributions and the covariance between different parts of the same photon history. As previously stated, the detailed physical interpretation of these covariance terms, including their dependence on the choice of history partition, will be addressed in a companion study (Kasiri et al., in prep). Here, we use this construction only to show that the moment-based convergence metrics are built from the full packet-level variance rather than from an artificially independent-event approximation.

\subsection{Zeroth-Order convergence metrics}

With the foundation set, the first step in assessing whether our simulations are converged is to examine the mean behaviour of the MC estimator. We refer to this as zeroth-order convergence. At this level, we are concerned only with the expected value of a physical quantity. We therefore define metrics for quantifying error and bias at zeroth order.

\subsubsection{Relative error of finite-sample estimates}

When dealing with a finite number of samples, it is crucial to understand the error that stems from limited sampling. The relative error diagnostic asks how quickly a finite photon group recovers the reference value measured from the full photon ensemble. It should not be interpreted as a physical bias. Instead, it measures finite-sample stability around the value selected by a given estimator.

We divide the full set of $N_\mathrm{ph}$ photons into $G$ disjoint groups, each containing $N$ photons. For group $j$, the group mean is
\begin{equation}
    \mu_j(N) = \frac{1}{N} \sum_{p=1}^{N} Y_{p,j} \, .
\end{equation}
Averaging over all groups gives the mean estimate at group size $N$,
\begin{equation}
    \bar\mu = \frac{1}{G} \sum_{j=1}^{G} \mu_j(N) \, .
\end{equation}
We estimate the reference value, $\mu_\mathrm{ref}$, using the full photon ensemble. The relative error of the mean is then
\begin{equation}
    \mathrm{RE}_{\mu}(N) = \frac{\bar\mu -\mu_\mathrm{ref}}
    {\mu_\mathrm{ref}} \, .
\end{equation}
An analogous quantity can be computed for the group-estimated standard deviation by replacing $\bar\mu$ with the standard deviation estimated from groups of size $N$.

Thus, $\mathrm{RE}=0$ means that the finite-group estimate agrees with the full-sample reference estimate. It does not mean that the estimator is physically unbiased relative to an analytic solution or trusted truth value. Physical bias is tested separately by comparing a converged estimator against an external benchmark. The relative error used here measures how rapidly the sample mean or sample standard deviation stabilises with photon count.

\subsubsection{Bias of the estimator}

As previously mentioned, bias for an estimator is detected via comparison against a chosen truth value. As a general definition,
\begin{equation} \label{eq:bias}
    \mathrm{Bias} \left( X \right) = \frac{X - X_\mathrm{truth}}{X_\mathrm{truth}}
    \, ,
\end{equation}
where $X$ represents a generic variable. The bias quantifies the difference between a given output and its expected value, with a value of 0 being ideal.

\subsection{First-order convergence metrics}

While zeroth-order convergence focuses on the accuracy of the estimator, first-order convergence addresses its precision. The key question is how precisely these quantities can be estimated for a given number of photons. In other words, we are concerned with the variance and size of statistical fluctuations around the mean.

\subsubsection{Coefficient of variation (CV)}

Before analysing the variance in an estimator, it is useful to understand variance at the per-photon level. To do this, we use the coefficient of variation (CV), a standard statistical measure of the dispersion of data around a mean. It is defined as
\begin{equation}
    \mathrm{CV} = \frac{\sigma}{\mu} \, .
\end{equation}
When applied to Ly$\alpha$ MCRT, CV is best understood as a measure of dispersion in the per-photon contribution distribution. A value of $\mathrm{CV} = 1$ means that the distribution is dispersed on the scale of its own mean. If $\mathrm{CV} \ll 1$, then the contributions fluctuate little and are tightly bound around the mean. If $\mathrm{CV} \gg 1$, then individual photon contributions fluctuate strongly.

\subsubsection{Fractional error (FE)}

To analyse the variance of our estimators, we begin by defining the fractional RMS error using the first and second raw moments introduced above:
\begin{equation}
    \mathrm{FE} \equiv \frac{\sqrt{\mathrm{Var}(S_1)}}{E[S_1]} = \frac{\sqrt{S_2 - \left( S_1^2/N_\mathrm{ph} \right)}}{S_1} \, .
\end{equation}
This can alternatively be equated as
\begin{equation}
    \mathrm{FE} = \frac{\sqrt{N_\mathrm{ph} \sigma_Y^2}}{S_1} = \frac{\sigma_Y}{\sqrt{N_\mathrm{ph}} \mu_Y} = \frac{\mathrm{CV}}{\sqrt{N_\mathrm{ph}}} \, .
\end{equation}
Fractional error is analogous to CV, but applied to the estimator level instead of the per-photon level. It measures the fractional uncertainty in the MC estimator after averaging over $N_\mathrm{ph}$ photons. If $\mathrm{FE} \ll 1$, then the final estimate is tightly bound to the true estimator mean. If $\mathrm{FE} \gg 1$, then the estimator is still dominated by sampling noise and is not trustworthy.

\subsubsection{Needed photons for desired variance}

With the fractional error metric established, the next step is to define how many photons would be needed for a desired level of uncertainty. We measure FE at a reference photon number $N_0$, and then infer the photon count required to achieve a target global fractional error $\mathrm{FE}_\mathrm{target}$ through
\begin{equation}
    N_\mathrm{needed} = N_0\left[\frac{\mathrm{FE}(N_0)}{\mathrm{FE}_\mathrm{target}}\right]^2 \, .
\end{equation}
Therefore, a minimum photon number can be extracted from a single run. This allows future runs to control finite-sampling error without using an unnecessarily costly number of photons.

\subsection{Second-order convergence metrics}

First-order metrics quantify variance at the photon and estimator levels, along with the photon count needed to reduce that noise. However, variance itself is an estimator built from a finite set of photons and therefore also fluctuates from realisation to realisation. Understanding how stable that variance estimate is requires pushing the convergence hierarchy one order further. At second order, the central question is how much the variance estimate would fluctuate if a simulation were rerun or repartitioned. The answer requires analysing the second and fourth moments of the underlying photon contribution distribution.

\subsubsection{Coefficient of variation of variance (CVoV)}

\begin{figure*}
    \centering
    \includegraphics[width=\textwidth]{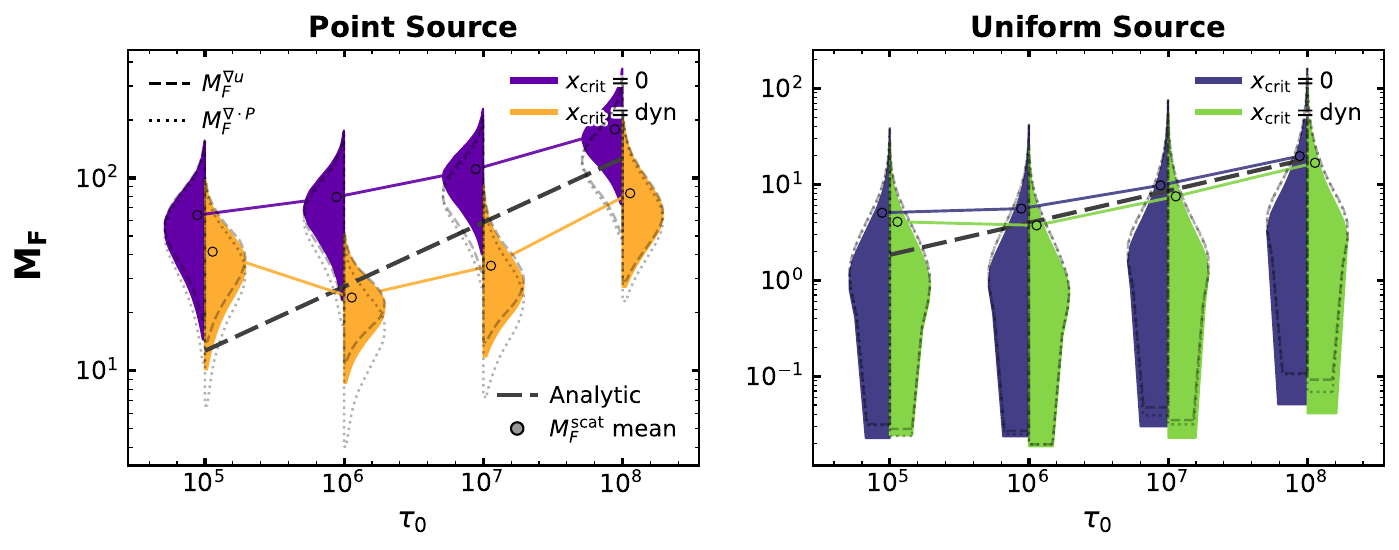}
    \caption{Integrated force multiplier distributions for central point-source and uniform-source emission as a function of line-centre optical depth. Violin distributions show the packet-level contributions to $M_\mathrm{F}$ for the direct scattering estimator, with colours indicating the no-core-skipping and dynamical core-skipping cases. Gray dashed and dotted curves show the corresponding path-based reconstructions from $\nabla u$ and $\nabla\cdot P$, while the black dashed curve gives the analytic diffusion-limit scaling. Filled circles mark the mean scattering-estimator force multiplier. The no-core-skipping scattering estimator provides the most direct reference for the finite-$\tau_0$ internal momentum deposition, while dynamical core skipping systematically lowers the integrated force. Path-based reconstructions follow the same broad trends, but show stronger estimator-level offsets for point-source transport than for uniform-source transport.}
    \label{fig:violin}
\end{figure*}

The primary second-order metric is the coefficient of variation of variance (CVoV). This is analogous to the usual coefficient of variation, but applied one level higher. Instead of measuring the dispersion of photon contributions $Y_p$, it measures the dispersion of finite-sample estimates of their variance.

Recall that $Y$ denotes the underlying per-photon contribution random variable, with independent realisations $Y_p$. The true packet-level variance is
\begin{equation}
\sigma_Y^2 \equiv \mathrm{Var}(Y) \, .
\end{equation}
In practice, however, $\sigma_Y^2$ is unknown and must be estimated from a finite sample of photon packets. We therefore divide the photon sample into $L$ disjoint groups, each containing $N$ photons. For group $j$, we define
\begin{equation}
\bar{Y}_j(N) = \frac{1}{N} \sum_{p=1}^{N} Y_{p,j}
\end{equation}
and the corresponding finite-sample variance estimate
\begin{equation}
\widehat{\sigma}_{Y,j}^{2}(N) = \frac{1}{N-1}
\sum_{p=1}^{N} \left[ Y_{p,j} - \bar{Y}_j(N) \right]^2 \, .
\end{equation}

In first-order convergence, this variance estimate is treated as a fixed estimate of the underlying packet-level variance $\sigma_Y^2$. It then determines the expected finite-sampling noise of the estimator through relations such as
\begin{equation}
\mathrm{Var}(S_1) = N_{\mathrm{ph}} \sigma_Y^2 \, ,
\end{equation}
and therefore enters the fractional error. In reality, however, $\widehat{\sigma}_{Y,j}^{2}(N)$ is itself a random quantity. A different group of $N$ photons would generally produce a different estimate of the same underlying variance. Second-order convergence quantifies the stability of these variance estimates.

The mean variance estimate across groups is
\begin{equation}
\mathrm{MoV}(N) = \frac{1}{L} \sum_{j=1}^{L} \widehat{\sigma}_{Y,j}^{2}(N) \, ,
\end{equation}
while the variance of the variance estimates is
\begin{equation}
\mathrm{VoV}(N) = \frac{1}{L-1} \sum_{j=1}^{L} \left[ \widehat{\sigma}_{Y,j}^{2}(N) - \mathrm{MoV}(N) \right]^2 \, .
\end{equation}
We then define the coefficient of variation of variance as
\begin{equation}
\mathrm{CoV}(N) = \frac{\sqrt{\mathrm{VoV}(N)}}{\mathrm{MoV}(N)} \, .
\end{equation}

Thus, $\mathrm{CVoV}$ measures the fractional uncertainty of the variance estimate itself. When $\mathrm{CVoV} \ll 1$, different photon subsamples give similar estimates of $\sigma_Y^2$, so the inferred FE is stable. When $\mathrm{CVoV} \gtrsim 1$, the variance estimate fluctuates by an amount comparable to its own mean, meaning that the quoted FE is not yet a robust characterization of the Monte Carlo uncertainty.

The behaviour of CVoV is governed by the fourth moment of the underlying packet-contribution distribution. Let $\mu_1=E[Y]$ be the mean packet contribution, $\sigma^2=E[(Y-\mu_1)^2]$ be the variance, and $\mu_4=E[(Y-\mu_1)^4]$ be the fourth central moment. We also use the kurtosis,
\begin{equation}
    \beta_2 = \frac{\mu_4}{\sigma^4} \, ,
\end{equation}
which measures the tail-heaviness of the distribution. For independent photon-packet contributions $Y_p$, classical large-sample statistics give an approximate expression for the variance of the subsample variance:
\begin{equation}
    \mathrm{Var}\left( \widehat{\sigma}_{Y,j}^2\right) \approx \frac{1}{N} \left(\mu_4 - \sigma^4\right) = \frac{\sigma^4}{N} \left(\beta_2 -1\right) \, .
\end{equation}
Taking the square root and normalizing by $E[s^2] \approx \sigma^2$, we obtain:
\begin{equation}
    \mathrm{CVoV} \approx \frac{\sqrt{\mathrm{Var}(s^2)}}{\sigma^2} \approx \frac{\sqrt{\beta_2 -1}}{\sqrt{N}} \, .
\end{equation}
Thus, CVoV scales as $N^{-1/2}$, with a prefactor determined by the kurtosis of the underlying distribution. For Gaussian distributions, $\beta_2 =3$ so CVoV $\approx \sqrt{2}/\sqrt{N}$. However, for heavy-tailed distributions, $\beta_2 > 3$, which leads to a larger CVoV, reflecting greater variability in variance estimates.

\section{Convergence of \texorpdfstring{Ly$\boldsymbol{\alpha}$ MCRT}{Convergence of Lya MCRT}}
\label{sec:mcrt_convergence}

With the foundations and metrics established, we now apply the hierarchy of convergence to the Monte Carlo estimators for Ly$\alpha$ momentum deposition. The goal is to determine at each level whether the estimator is converged. The three key questions are whether the mean is biased, how much noise arises from finite sampling, and whether the variance estimate itself is stable. These questions correspond to zeroth-, first-, and second-order convergence, respectively. Throughout this section, each diagnostic is applied to the same set of source geometries, physical parameters, and estimator constructions so the competing methods can be compared directly.

The most direct estimator is the event-based scattering estimator. It accumulates contributions at each individual scatter and most closely follows the physical momentum exchange. We write the dimensionless radial momentum transfer schematically as
\begin{equation} \label{eq:S_scat}
    S_\mathrm{scat} = \sum_e w
    \left[ \left( \bm{n}_\mathrm{old} - \bm{n}_\mathrm{new} \right) \cdot \hat{\bm{r}} \right] \, .
\end{equation}
Here $w$ is the packet weight, normalised to unity across all photons. Acceleration, force, and force multiplier all share this same underlying estimator, but receive different global or cell normalisations. These three quantities can therefore be discussed somewhat interchangeably when considering MC noise, because they are generated by the same packet-level random variable. In particular,
\begin{equation}
    M_\mathrm{F}^\mathrm{scat} = S_\mathrm{scat} \, ,\qquad F^\mathrm{scat} = \frac{\mathcal{L}}{c}S_\mathrm{scat}  \, , \qquad a^\mathrm{scat} = \frac{\mathcal{L}}{c \rho V} S_\mathrm{scat} \, .
\end{equation}
Thus, the key distinction between acceleration, force, and force multiplier is how they are rescaled at the end of the simulation. They do not differ in the photon histories they sample.

We also consider path-based constructions of the force. These do not tally the momentum kicks directly. Instead, they estimate moments of the radiation field through photon path lengths and then reconstruct the force with a spatial derivative. The energy density estimator is
\begin{equation}
    u = \frac{\mathcal{L}}{c V} \sum_{\mathrm{paths}} w \Delta \ell \, ,
\end{equation}
which gives the diffusion-limit force construction
\begin{equation}
    a^{\nabla u} \equiv -\frac{1}{3\rho} \nabla u \, .
\end{equation}
Similarly, the radiation-pressure estimator is
\begin{equation} \label{eq:pressure-const}
    P = \frac{\mathcal{L}}{c V} \sum_{\mathrm{paths}} w \int_{\mathrm{path}} \mu^2 d\ell \, ,
\end{equation}
where $\mu \equiv \bm{n} \cdot \hat{\bm{r}}$. This then reconstructs the acceleration through
\begin{equation}
    a^{\nabla \cdot P} = -\frac{1}{\rho} \left( \nabla \cdot P \right) \, .
\end{equation}
In practice, the path integral for Eq.~(\ref{eq:pressure-const}) is evaluated analytically rather than by quadrature. For a straight path segment with impact parameter $b$, let $s$ denote the signed distance along the ray measured from closest approach. Then $r^2 = s^2 + b^2$ and $\mu = s/r$, giving
\begin{equation}
    \int_{\ell_1}^{\ell_2}\mu^2\,d\ell = \left[ s-b\tan^{-1}\left(\frac{s}{b}\right) \right]_{s_1}^{s_2} \, .
\end{equation}
This expression assumes that the packet weight is constant along the path segment, and is therefore exact for the dust-free calculations considered here. In the presence of continuous absorption, the weight would vary along the segment and the expression would need to be modified.

\begin{figure*}
    \centering
    \includegraphics[width=\textwidth]{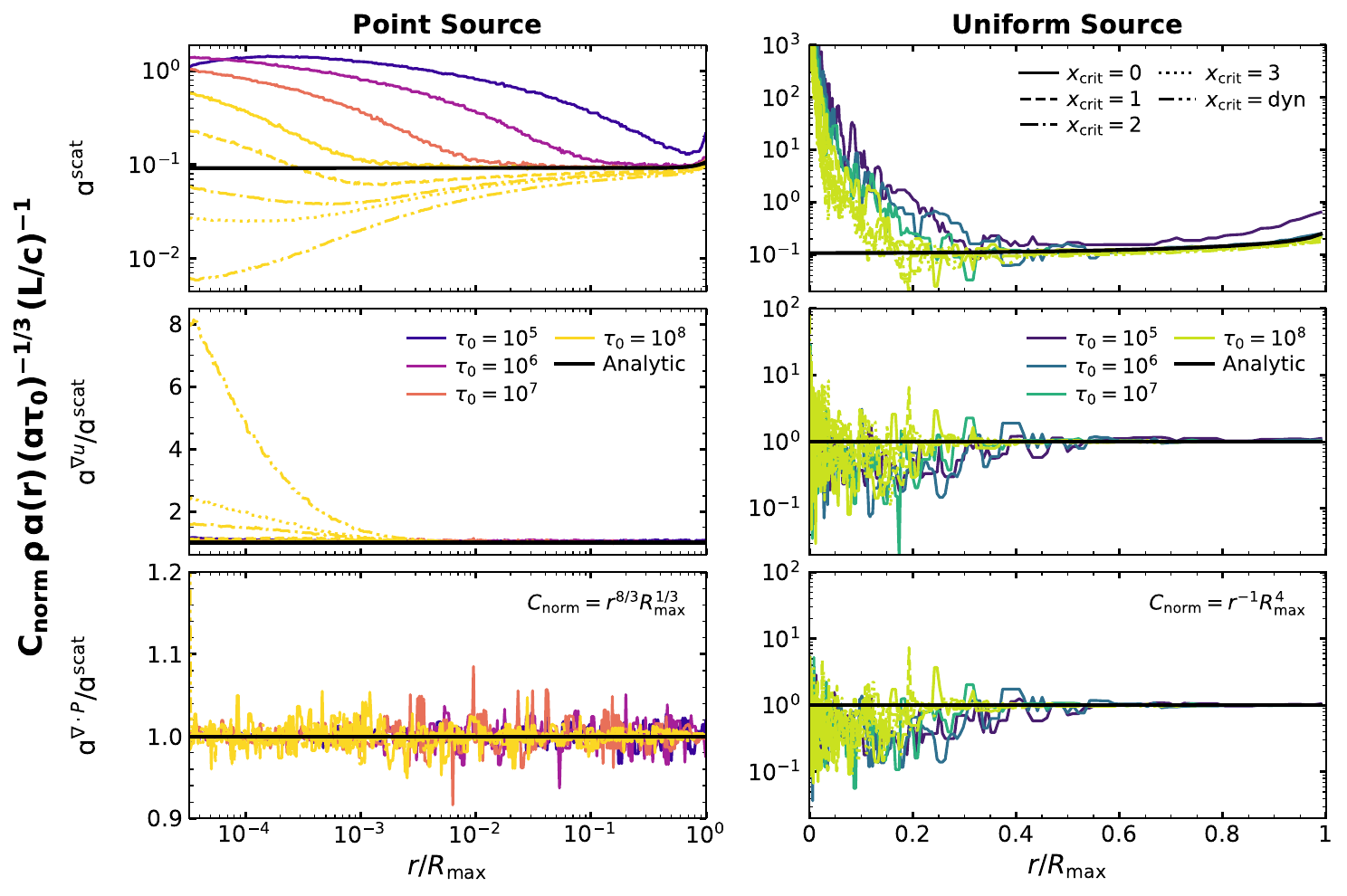}
    \caption{Radial acceleration profiles for point-source and uniform-source emission. The top row shows the direct scattering estimator $a^\mathrm{scat}$ and the bottom two rows show the gradient-of-energy-density $a^{\nabla u}$ and radiation-pressure-divergence $a^{\nabla\cdot P}$ reconstruction relative to the direct scattering case. Colours indicate line-centre optical depth, while line styles indicate the core-skipping prescription $x_\mathrm{crit}$. Black curves show the analytic diffusion-limit profiles. Profiles are scaled by $(a\tau_0)^{-1/3}(\mathcal{L}/c)^{-1}$ and by the panel-dependent normalisation $C_\mathrm{norm}$ indicated in the figure. For point sources, increasing core skipping suppresses the internal acceleration, especially in the inner cloud. For uniform sources, the profiles are broadly consistent outside the geometric core, while the innermost region is noisy because unbiased volume-weighted launching provides few central photon histories.}
    \label{fig:accel-profile}
\end{figure*}

\begin{figure*}
    \centering
    \includegraphics[width=\textwidth]{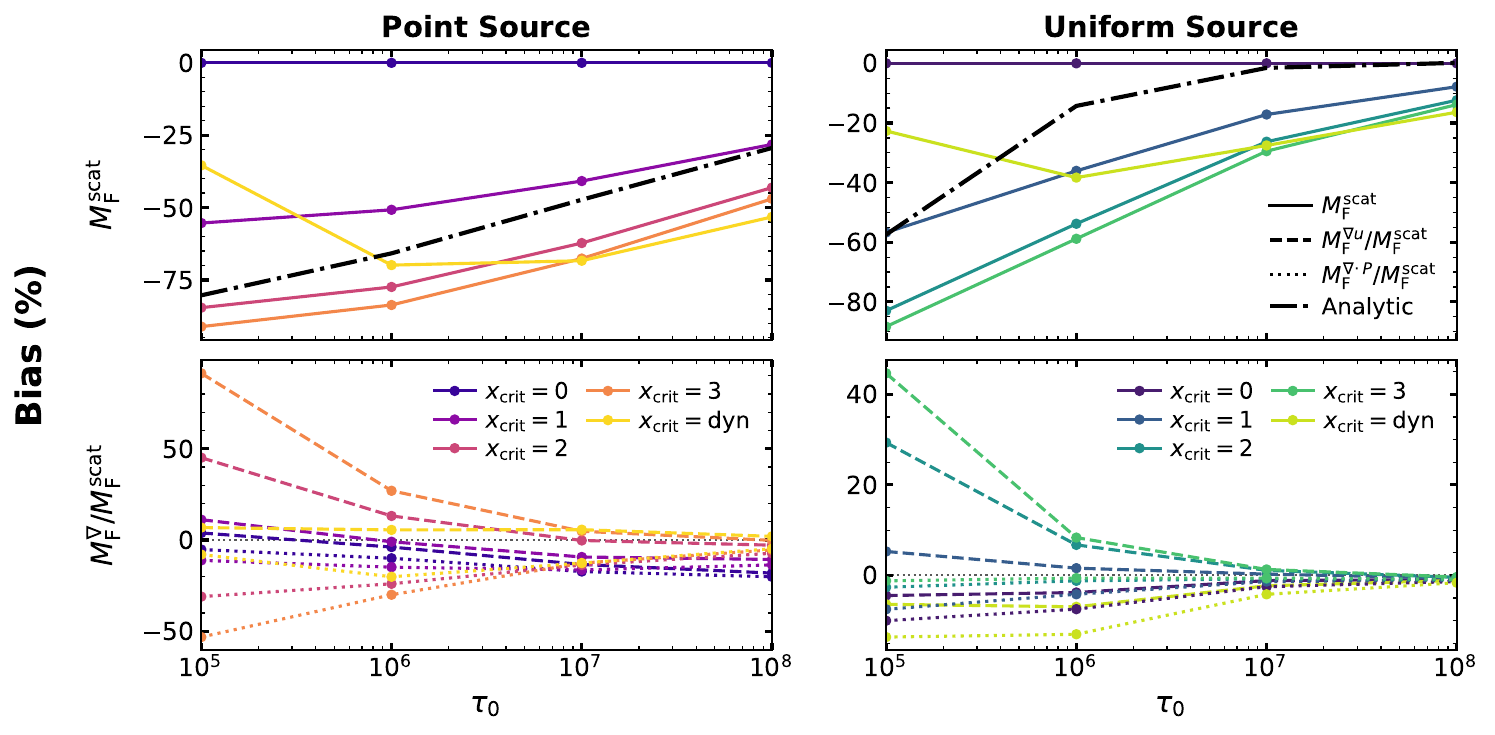}
    \caption{Integrated force-multiplier bias as a function of optical depth for the three force constructions. The top row bias is computed relative to the no-core-skipping direct scattering estimator at each optical depth, so the $x_\mathrm{crit}=0$ scattering curve is zero by construction. The bottom row shows $M_\mathrm{F}^{\nabla u}$ and $M_\mathrm{F}^{\nabla\cdot P}$ relative to $M_\mathrm{F}^\mathrm{scat}$ of the corresponding optical depth and core skipping prescription. Columns show point-source and uniform-source emission. Coloured curves indicate fixed and dynamical core-skipping prescriptions, and the black dash-dotted curve shows the analytic diffusion-limit comparison. Core skipping generally produces negative bias by omitting scatterings that contribute to the internal momentum budget. The path-based estimators show additional reconstruction bias for point-source emission, while for uniform-source emission they more closely track the scattering-estimator trends.}
    \label{fig:bias}
\end{figure*}

Although we define path-based force estimators through $\nabla u$ and $\nabla \cdot P$, we find that they are less robust than the direct scattering estimator of Eq.~(\ref{eq:S_scat}) in the extremely optically thick regime. We also expect these estimators to show different convergence behaviour from the direct event-based estimator because their noise is controlled by path-length sampling and spatial reconstruction instead of the usual discrete sequence of kicks. The spatial derivatives in the path-based reconstructions are evaluated using the finite-difference and finite-volume stencils described in Appendix~\ref{app:spatial_stencils}.

These three estimators are tested for both central point-source and uniform-source emission in static spherical clouds. The two source types have distinct numerical difficulties. For uniform-source emission, photons are launched proportional to volume. This creates an undersampling issue within the geometric core, where too few photon histories are launched to yield meaningful statistics. For point-source emission, the radiation field is centrally concentrated. Force reconstructions are therefore sensitive to the spatial gradient and to the assumptions entering the diffusion or closure relation. These source-dependent caveats should be kept in mind when interpreting the results below.

Unless otherwise stated, all simulations were performed with a photon count of $N_\mathrm{ph}=10^6$, a gas temperature of $T=10^4$ K, and 500 radial and frequency bins. Optical-depth parameter sweeps covered the range $\tau_0=10^5$ to $\tau_0=10^8$, which, for the adopted temperature, corresponds to $a\tau_0 \simeq 47.0$ to $4.70\times10^4$, using $a=4.702\times10^{-4}(T/10^4 \, \mathrm{K})^{-1/2}$. The default core-skipping prescription is none, although cases with static values of 1, 2, and 3, as well as the dynamical prescription, are considered where specified. Throughout this study, the gas geometry is assumed to be spherical, and the effects of velocity gradients and dust are neglected.

\subsection{Zeroth-order convergence: mean force, profiles, and bias}

The most basic question for convergence is whether the estimator has reached the correct mean. This is distinct from whether the estimator has small variance. A low-noise estimator can converge to an incorrect value, while a noisy one can remain unbiased in expectation value.

\subsubsection{Per-photon contribution distributions}

Fig.~\ref{fig:violin} shows this distinction at the level of the integrated force multiplier per photon. For point-source emission, the no-core-skipping event-based estimator provides the most direct numerical measurement of the finite $\tau_0$ momentum deposited in the gas. Its mean increases with optical depth and remains close to the analytic scaling at large $\tau_0$. A modest overestimate remains even at the highest opacity, although Fig.~\ref{fig:validation} shows little visible profile discrepancy in this regime. This apparent difference reflects what each diagnostic weights. Fig.~\ref{fig:validation} is primarily a local, logarithmic test of the radial profile shape, whereas $M_\mathrm{F}$ is a shell-weighted integral. Therefore, small coherent offsets that are visually minor in $a(r)$, especially near the radii that dominate the shell-weighted contribution or near the diffusion-to-escape transition, can accumulate into a measurable offset in the integrated force multiplier. The width of the no-core-skipping $M_\mathrm{F}$ distribution also tightens as the medium becomes more opaque. Even so, these distributions remain broad, with individual photon realisations spanning a wide range of contributions. Dynamical core skipping produces systematically smaller force multipliers because the algorithm skips events that would otherwise contribute to the estimator. The violin plot already shows that core skipping both lowers the variance and biases the resulting internal forces.

The path-based force multiplier reconstructions follow the general shape of the event-based distributions. However, their agreement with the scattering estimator depends on source geometry. In the point-source case, both the gradient-of-energy-density and divergence-of-pressure reconstructions show an offset from the no-core-skipping scattering curves. This becomes most pronounced at the highest optical depths. Dynamical core skipping behaves better at high optical depth for the reconstruction methods. This offset arises because the path-based construction converts a radiation-field moment into a force using a finite-resolution spatial derivative and an assumed boundary-condition closure relation. These operations are most sensitive for centrally concentrated point-like sources with steep transport gradients. In the uniform-source case, the path-based and scattering-based integrated force multipliers agree more closely, consistent with the smoother radiation field and absence of a central singularity from the source function.

\subsubsection{Acceleration radial profiles}

The radial acceleration profiles in Fig.~\ref{fig:accel-profile} illustrate where the integrated trends originate. For the point source, the no-core-skipping profiles retain the full momentum deposition and provide the comparison standard. As core-skipping strength increases, the curves increasingly underestimate the force. The bias is most evident in the geometric core. This suppression occurs for all three construction methods, indicated by the near order-unity comparison values. Core skipping therefore does more than affect the discrete scattering tally; it alters the internal radiation field. The analytic profile also agrees well with the no-core-skipping case once the cloud is optically thick enough. At lower optical depth, the shape and magnitude of the profile stray from the analytic solution because the calculation moves out of the regime where the analytic solution is valid.

The uniform-source profiles show a different limitation. The profiles approach a unified behaviour through much of the cloud, but the central spatial region is noisy and irregular. This is the geometric-centre undersampling issue (see Appendix~\ref{app:uniform_volume_boosting}). In unbiased uniform launching, only a small fraction of photons are born at very small radii. The inner bins therefore receive few independent source histories, making the local acceleration difficult to estimate. Outside this central region, the curves are much more stable, and the separation between different optical depths and core-skipping prescriptions is less dramatic than in the point-source case.

\subsubsection{Integrated force multiplier bias} \label{sec:bias}

\begin{figure*}
    \centering
    \includegraphics[width=\textwidth]{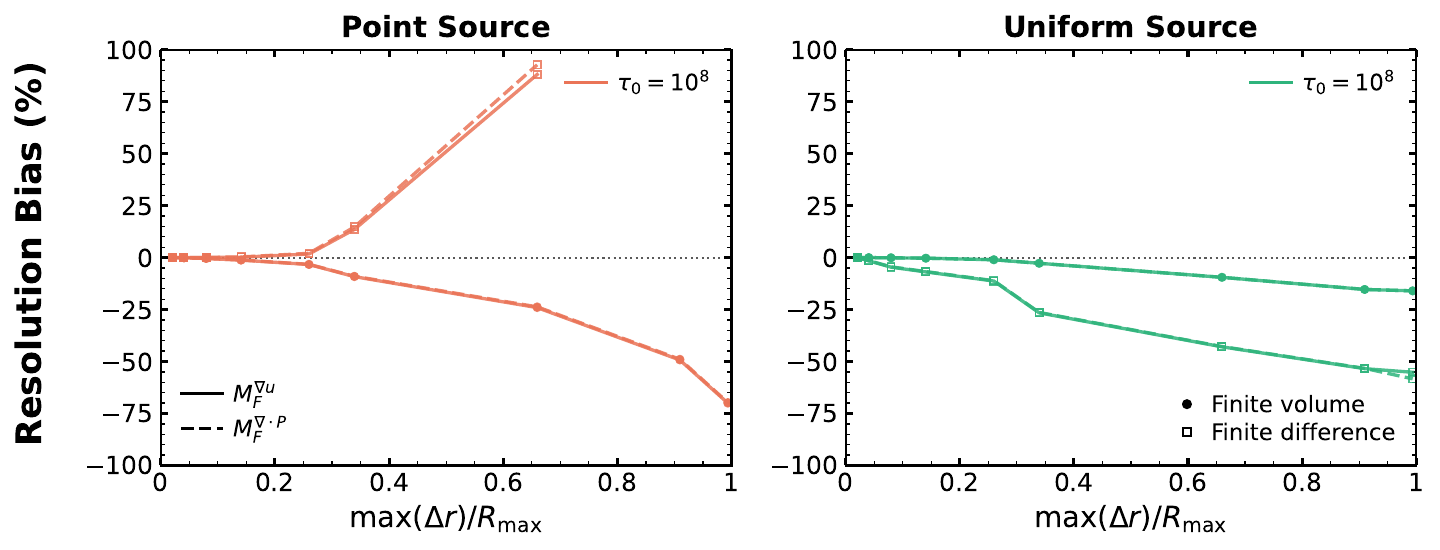}
    \caption{Resolution bias of the path-based force reconstructions at $\tau_0=10^8$. The horizontal axis gives the maximum radial cell width normalised by the cloud radius (near 0 represents native bin count and 1 represents single bin count), and the vertical axis gives the integrated force-multiplier bias relative to the high-resolution reference calculation. Curves with filled circles show the finite-volume stencil, while curves with open squares show the finite-difference stencil. Solid and dashed line types distinguish the $\nabla u$ and $\nabla\cdot P$ reconstructions. Coarsening the radial grid strongly affects the point-source force reconstruction because the central radiation-field gradient is poorly resolved. The uniform-source case is more robust, but still develops non-negligible bias at coarse resolution. The two derivative stencils are described in Appendix~\ref{app:spatial_stencils}.}
    \label{fig:res-bias}
\end{figure*}

The integrated bias in Fig.~\ref{fig:bias} compresses these profile-level effects into a single zeroth-order diagnostic. We use the bias definition in Eq.~(\ref{eq:bias}), setting the truth value in the top row to the no-core-skipping event-based estimator at each optical depth. Therefore, for both sources, the no-core-skipping scattering runs are zero by construction. Nonzero core-skipping prescriptions produce negative bias, confirming that skipped core scatterings remove real internal momentum deposition. The bias generally decreases in magnitude as optical depth increases, likely because the signal becomes increasingly dominated by rare excursion events rather than contributions accrued through core scatterings, making skipped core scatterings less consequential. However, the bias remains significant for aggressive core skipping. The dynamical prescription is less uniformly biased than fixed large $x_\mathrm{crit}$ choices, but it still does not reproduce the no-core-skipping force exactly.

The bottom row shows the path-based estimator bias relative to the corresponding event-based estimator. It does not compare specifically to the no-core-skipping prescription; it compares to the matching $x_\mathrm{crit}$ value. Therefore, a bias of $0\%$ for $M_\mathrm{F}^{\nabla} / M_\mathrm{F}^\mathrm{scat}$ only indicates the absence of additional estimator-specific bias. This reveals an additional distinction between source geometries. For point sources, $M_\mathrm{F}^{\nabla u}$ and $M_\mathrm{F}^{\nabla \cdot P}$ are biased even when no core scatterings are skipped. This indicates an estimator-level reconstruction error on top of any core-skipping bias. The path-based estimators infer a force from spatial derivatives of radiation moments. In a point-source problem, the steep central structure and core-wing transition make that inference imperfect. In contrast, for uniform sources the path-based curves collapse onto zero bias relative to the scattering estimator. The remaining bias is then dominated by the physical effect of core skipping rather than by a disagreement between estimator families.

\begin{figure*}
    \centering
    \includegraphics[width=\textwidth]{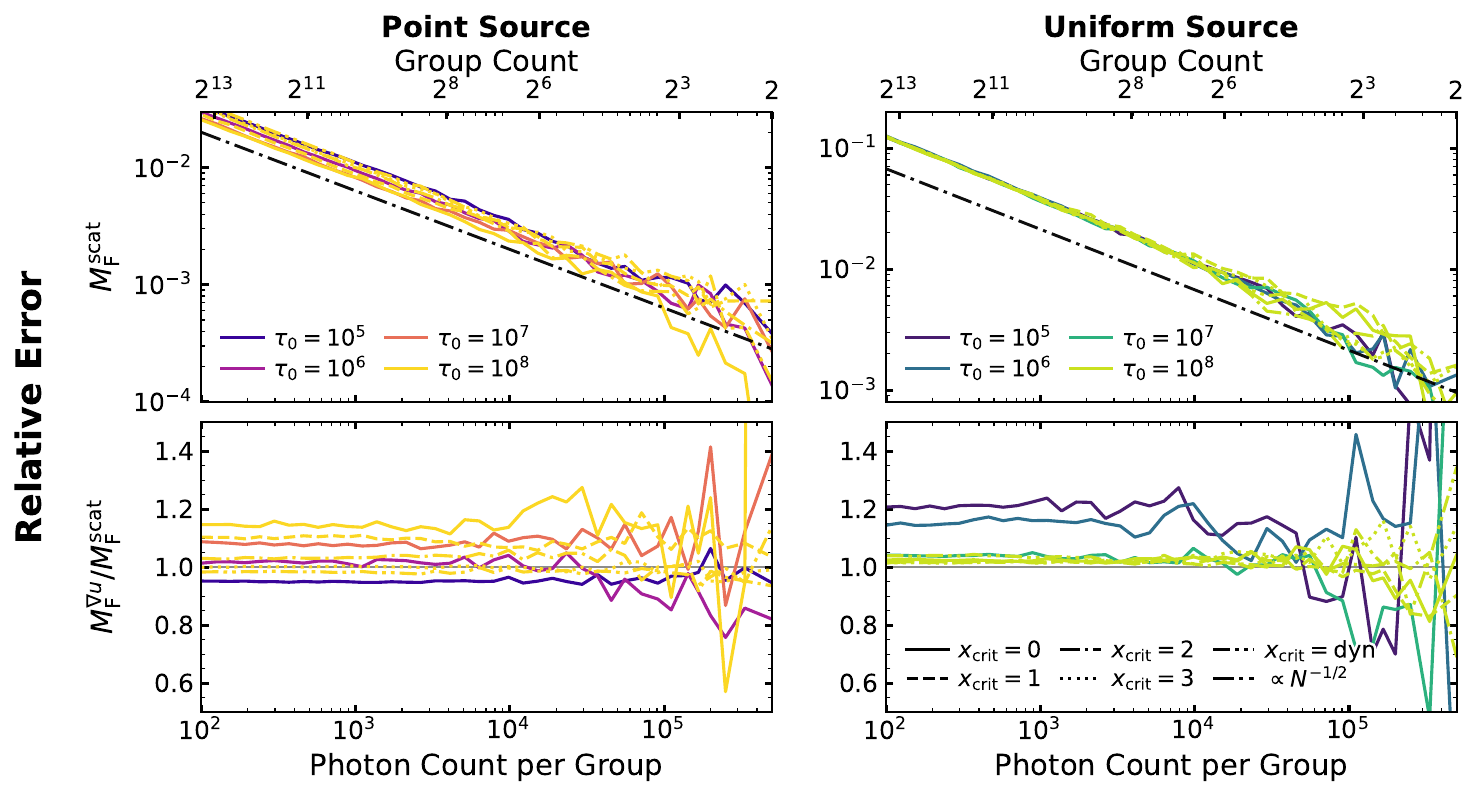}
    \caption{Finite-group relative error as a function of photon count per group. Top row shows the direct scattering estimator, while the bottom row displays the gradient-of-energy-density reconstruction relative to scattering estimator; columns show point-source and uniform-source emission. The lower axis gives the photon count in each group, while the upper axis gives the corresponding number of independent groups. Colours indicate optical depth and line styles indicate the core-skipping prescription. The black dash-dotted guide shows the expected $N^{-1/2}$ Monte Carlo scaling. The relative error measures how quickly finite photon groups recover the full-sample value of the same estimator, and should therefore be interpreted as a self-convergence diagnostic rather than as a physical bias. All constructions follow the expected scaling over most of the sampled range, with scatter at the largest group sizes caused by the small number of remaining independent groups.}
    \label{fig:relerr}
\end{figure*}

\subsubsection{Resolution bias of force reconstruction}

\begin{figure*}
    \centering
    \includegraphics[width=\textwidth]{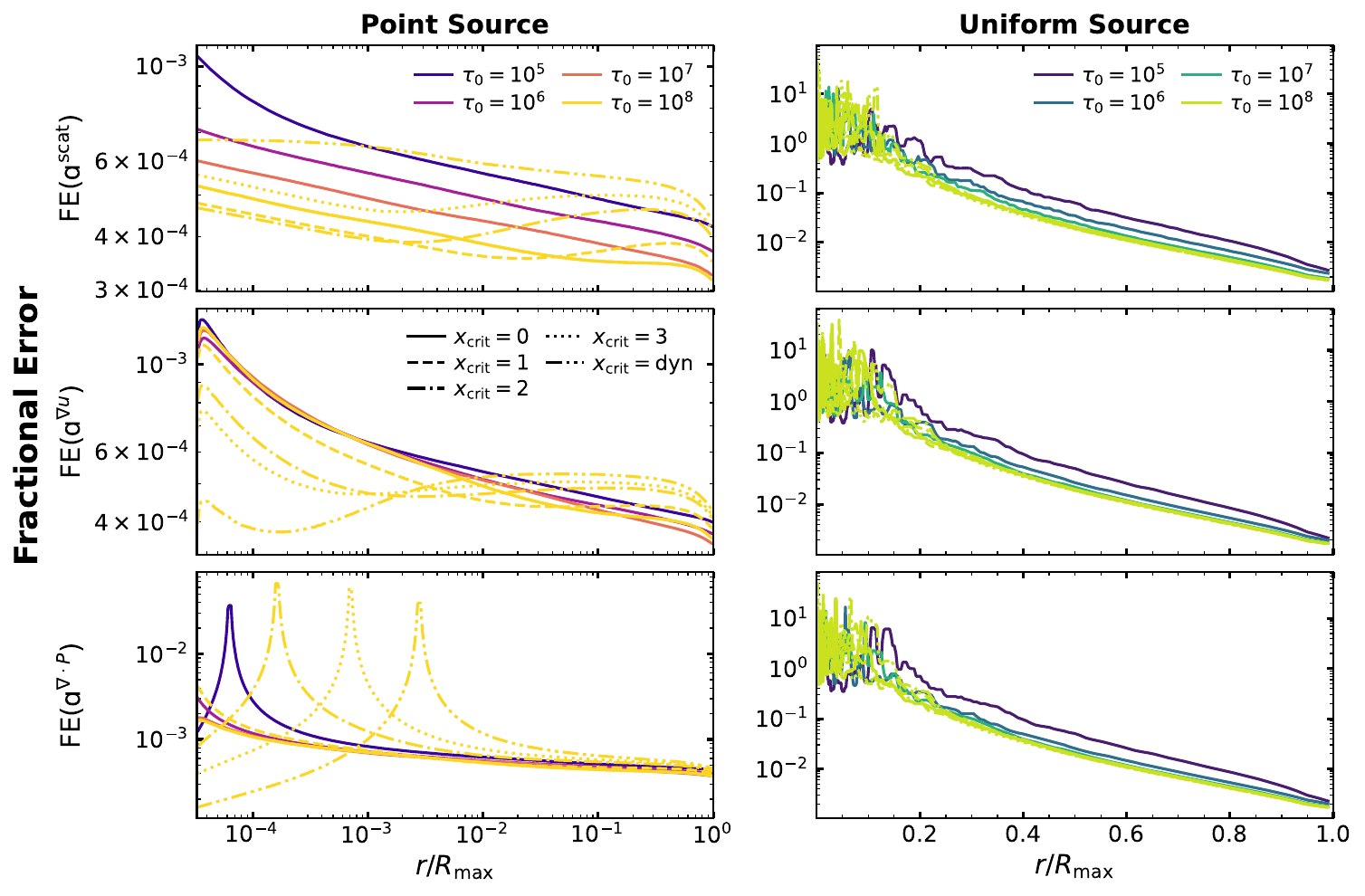}
    \caption{Radial profiles of the fractional error for the three force constructions. Rows show $\mathrm{FE}(a^\mathrm{scat})$, $\mathrm{FE}(a^{\nabla u})$, and $\mathrm{FE}(a^{\nabla\cdot P})$; columns show point-source and uniform-source emission. Colours indicate optical depth and line styles indicate the core-skipping prescription. For point-source emission, the fractional errors are small across most of the domain, with localised features appearing most clearly in the $\nabla\cdot P$ reconstruction. For uniform-source emission, the innermost region has much larger and noisier fractional error because unbiased uniform launching undersamples the small-volume geometric core. The cumulative fractional error decreases toward large radius as more photon histories contribute to the integrated force.}
    \label{fig:fe}
\end{figure*}

Resolution tests in Fig.~\ref{fig:res-bias} further support this interpretation. In the point-source case, coarsening the radial grid strongly amplifies the bias in the path-based force reconstruction. This is especially true for the finite-difference stencil, for which the bias can exceed order unity. For the uniform-source case, the two stencils behave in a far more stable manner, although both still develop non-negligible bias as the grid is coarsened. The two path-based constructions never diverge strongly from one another as resolution is degraded. This is expected because the underlying path-length tallies are unchanged, so the differences isolate the deterministic reconstruction step; the two derivative stencils are defined in Appendix~\ref{app:spatial_stencils}. Once the spatial gradient is smeared over too large a radial cell, the inferred force no longer faithfully represents the direct momentum deposition. The uniform case is more robust because its radiation field is smoother. Thus, the path-based estimators carry a spatial-resolution requirement that is more severe for point-source emission and depends on the stencil.

\subsubsection{Relative error of finite-sampling}

The last zeroth-order metric is the relative error, shown in Fig.~\ref{fig:relerr}. This metric does not test the physical correctness of a quantity and should not be confused with bias. Instead, it quantifies how quickly group estimates approach the full-sample estimator value. It asks how many photons are needed to recover the result obtained from the full photon ensemble. The gradient-of-energy-density construction is shown relative to the scattering construction. A value near order unity means that the behaviour of the scattering construction is shared by the gradient-of-energy-density construction. $M_\mathrm{F}^{\nabla \cdot P}$ was omitted because it closely resembled the $M_\mathrm{F}^{\nabla u}$ trends. For all source geometries and construction methods, a $N^{-1/2}$ scaling is present, as expected when the mean averages over independent photon samples. For the point source, it takes $\sim 10^3$ photons to reach $1 \%$ relative error in the group mean. For the uniform source, it takes $\sim 10^4$ photons to reach the same level. Path-based reconstructions have slightly higher relative error at high optical depth for point-source emission. For uniform-source emission, the opposite is true, with lower optical depths showing higher relative error. In either case, the comparative values remain near order unity. Therefore, expensive photon counts are not required to recover each estimator's respective full-sample mean for the simulation parameters considered here. Deviations at the largest group sizes are expected because the number of independent groups becomes too small for reliable statistics.

\subsection{First-order convergence: sampling noise and photon cost}

With mean behaviour addressed, we now turn to the variance of the estimator. Even if a mean appears physically correct, it may still have high noise. A quantity converged in zeroth order is therefore not necessarily converged in first order. The variance must be addressed separately.

\subsubsection{Fractional error of the force}

The spatial structure of the first-order error is shown in Fig.~\ref{fig:fe}. For point-source emission, the fractional errors are small across most of the domain. The scattering and $\nabla u$ constructions are especially stable, remaining within similar FE ranges across all optical depths and core-skipping prescriptions. However, the $\nabla \cdot P$ construction shows localised features associated with reconstructing a derivative of a higher angular moment. These features are most prevalent at low optical depths or for strong core-skipping prescriptions, and wash out once enough volume is included. Even in these cases, the point-source fractional errors remain modest. This means that, once the estimator target value is fixed, relatively few photons are needed to estimate that target precisely.

The uniform-source FE profiles are qualitatively different. The inner region shows large and irregular fractional errors, sometimes exceeding order unity. These then decrease smoothly toward the outer radius. This confirms that the uniform-source difficulty is primarily a spatial sampling issue concentrated in the geometric core, rather than an integrated normalisation issue. As more of the cloud volume is included, the cumulative estimate stabilises, and the final integrated FE becomes much smaller than the central-bin values. Nevertheless, the uniform-source calculation remains more variance-limited than the point-source calculation at the same total photon count.

\begin{figure*}
    \centering
    \includegraphics[width=\textwidth]{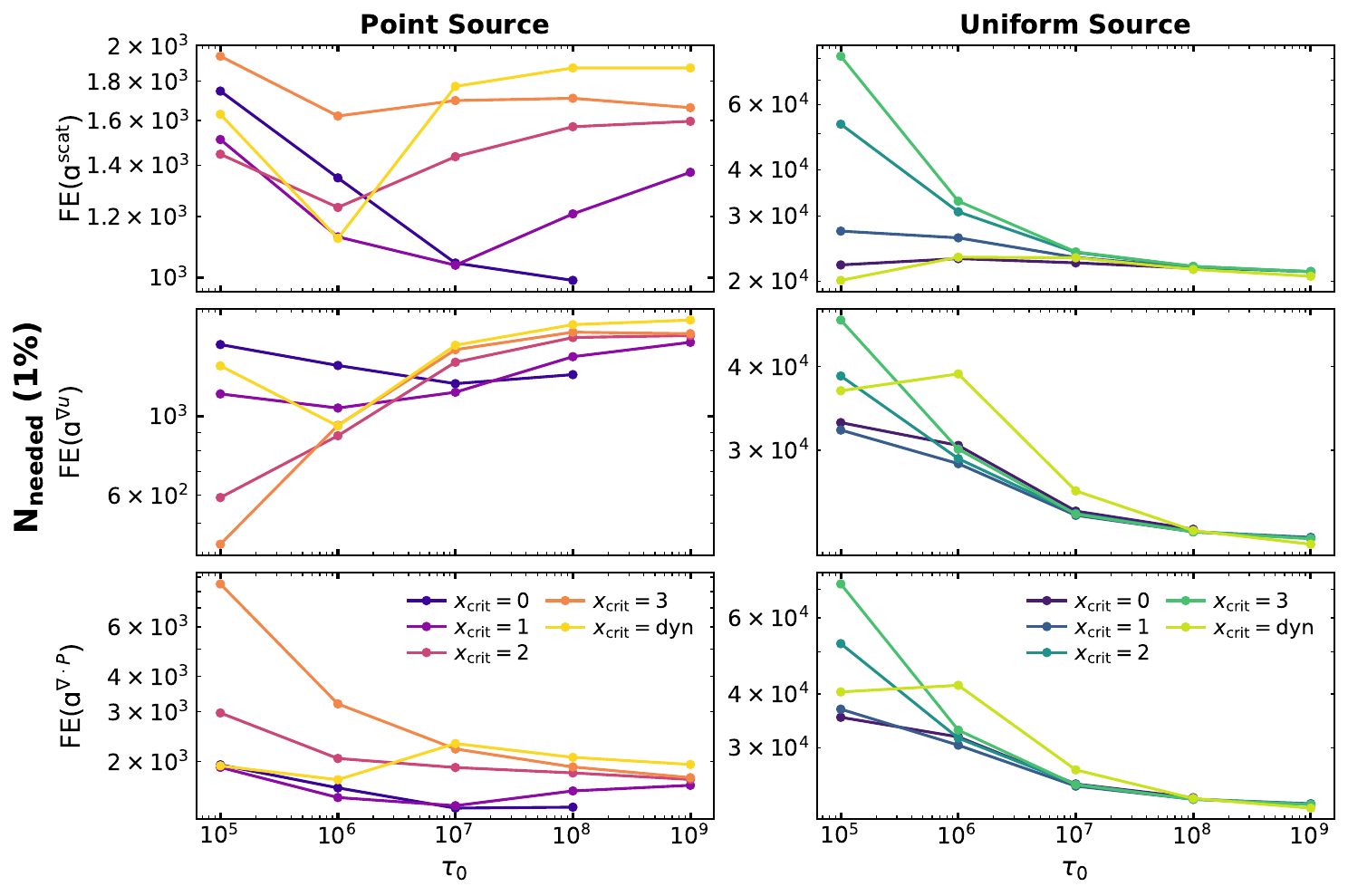}
    \caption{Photon number required to reach a target integrated fractional error of $1\%$. Rows correspond to the three force constructions, and columns show point-source and uniform-source emission. The required photon count is inferred from the measured fractional error using the expected $N^{-1/2}$ Monte Carlo scaling. Point-source calculations typically require only $\sim 10^3$--$10^4$ photons for the integrated force metrics shown here, while uniform-source calculations require $\sim 10^4$--$10^5$ photons because the geometric core is sampled less efficiently. These values quantify statistical precision around each estimator's own mean; they do not by themselves test whether that mean is physically unbiased.}
    \label{fig:n-needed}
\end{figure*}

It may seem surprising that all force constructions have the same order of magnitude in FE. The usual intuition is that the direct scattering estimator should have much higher variance, while the path-based estimators should have much lower variance. This picture is likely correct for the raw variance that would arise from independent contributions. However, the final estimator variance also depends on covariance. As noted above, a detailed breakdown of how covariance enters the estimator will be presented in a companion paper (Kasiri et al., in prep). A brief interpretation is that, for the event-based case, covariance suppresses the total variance by many orders of magnitude because momentum kicks can telescope. For the two path-based cases, covariance instead reinforces the total variance by an order of magnitude or two. The intuition is that path length is a positive, contiguous residence-time measure. Long-lived photons tend to over-contribute across neighbouring cells coherently, so their fluctuations reinforce rather than telescope away. This brings the estimators into the same broad variance range. The predicted scaling in Section~\ref{sec:momentum_scaling} is also realised in the data. The absence of a clear optical-depth dependence further supports this heuristic explanation.

\subsubsection{Photons needed for target noise}

Fig.~\ref{fig:n-needed} converts these fractional errors into the photon count required to reach a target precision of one percent. For point-source calculations, the required photon numbers are typically only $\sim 10^3$ to $10^4$ for the integrated force metrics shown here. The dependence on optical depth and core-skipping strength is relatively weak compared with the source-geometry dependence. For uniform sources, the required photon counts rise to the $\sim 10^4$ to $10^5$ range, reflecting the additional cost of sampling the inner source volume. At high optical depth, the different core-skipping choices tend to cluster more closely, while at lower optical depth the sampling cost is more sensitive to estimator and core-skipping prescription. This can be interpreted as higher optical depth producing stronger trapping episodes in the frequency core, and therefore in the geometric core where these photons are undersampled. Trapping in this region partly counteracts the lack of launch statistics, because each photon contributes more before travelling to the outer regions of the gas cloud.

These photon requirements should be interpreted as precision requirements, not accuracy requirements. They tell us how many photons are needed to reduce finite-sampling fluctuations around the estimator's own mean. They do not guarantee that the estimator mean is physically correct. The plot can also be misleading because similar required photon counts across core-skipping prescriptions and optical depths do not imply similar runtime costs.

\subsubsection{Time needed for target noise}

\begin{figure*}
    \centering
    \includegraphics[width=\textwidth]{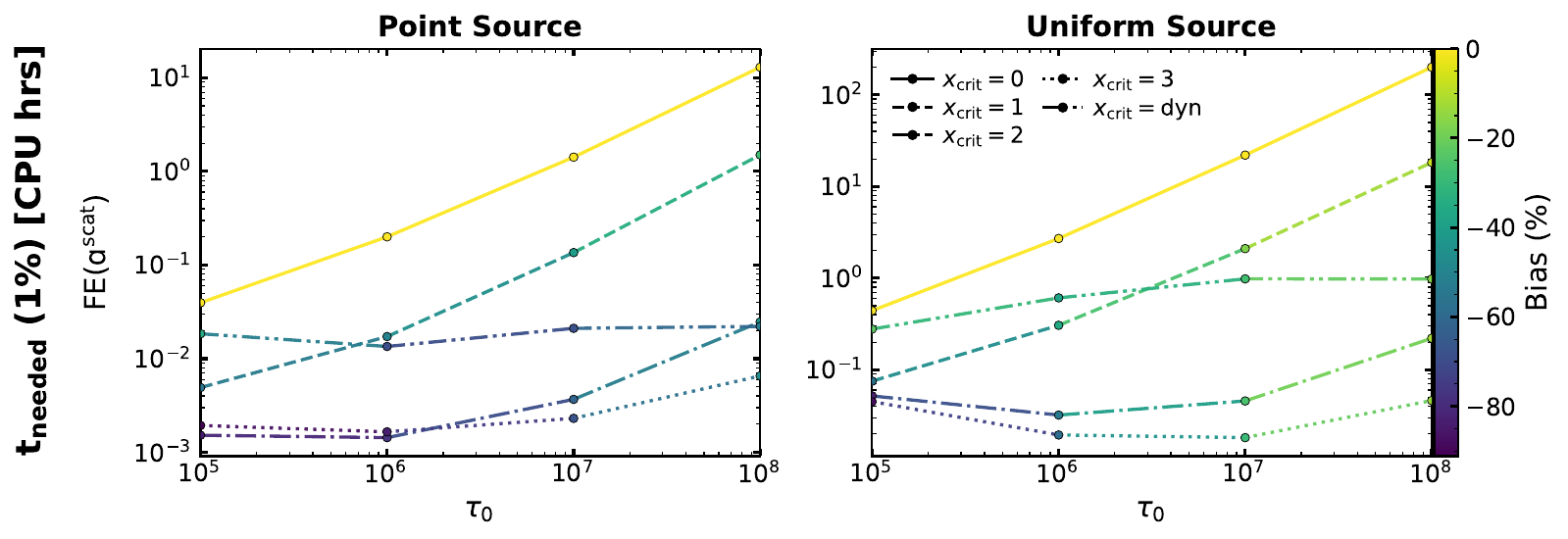}
    \caption{Estimated normalised runtime required to reach a target integrated fractional error of $1\%$ for the direct scattering estimator. The required photon number is converted to a wall-clock time using the measured time per photon for each run and then multiplied by the number of ranks and threads utilized. Colours show the integrated force-multiplier bias of each core-skipping prescription, and line styles indicate $x_\mathrm{crit}$. Core-skipping prescriptions reduce the computational cost per target precision, especially at large optical depth, but this speedup is accompanied by negative force bias. Dynamical core skipping has a weaker optical-depth dependence in runtime than the no-core-skipping calculation, illustrating the tradeoff between computational efficiency and physical fidelity for internal force estimates.}
    \label{fig:t-needed}
\end{figure*}

Fig.~\ref{fig:t-needed} extends the previous plot to computational cost. To estimate the cost differences, the time per photon was calculated for every run and then multiplied by the number of ranks and threads used. This normalised runtime per photon was then used to convert $N_\mathrm{needed}$ to $t_\mathrm{needed}$. The bias discussed in Section~\ref{sec:bias} is also included. The main difference between source geometries is that the uniform-source runtime is about an order of magnitude higher, as in the $N_\mathrm{needed}$ plot. For both sources, the no- and low-core-skipping prescriptions require the longest runtime and have the steepest scaling with optical depth. Between $\tau_0 = 10^7$ and $\tau_0 = 10^8$, the no-core-skipping prescription scales as $\propto \tau_0^{0.961}$ for point-source emission and $\propto \tau_0^{0.959}$ for uniform-source emission. More aggressive core skipping generally requires fewer photons and has weaker scaling with cloud opacity. Dynamical core skipping is nearly invariant to optical depth in all scenarios. However, as already discussed, core-skipping algorithms introduce negative bias. They make the runs more computationally efficient but less physically accurate. The two force reconstruction methods were omitted because their timing plots were nearly identical to the scattering-based case.

\subsection{Second-order convergence: stability of the variance estimate}

The first-order diagnostics depend on an estimated variance, which itself can vary from one realisation to another. We therefore need to test the variance of the variance estimator itself. If the variance estimator is stable, then conclusions drawn from one instance will be applicable across different instances.

\subsubsection{Coefficient of variation of variance}

\begin{figure*}
    \centering
    \includegraphics[width=\textwidth]{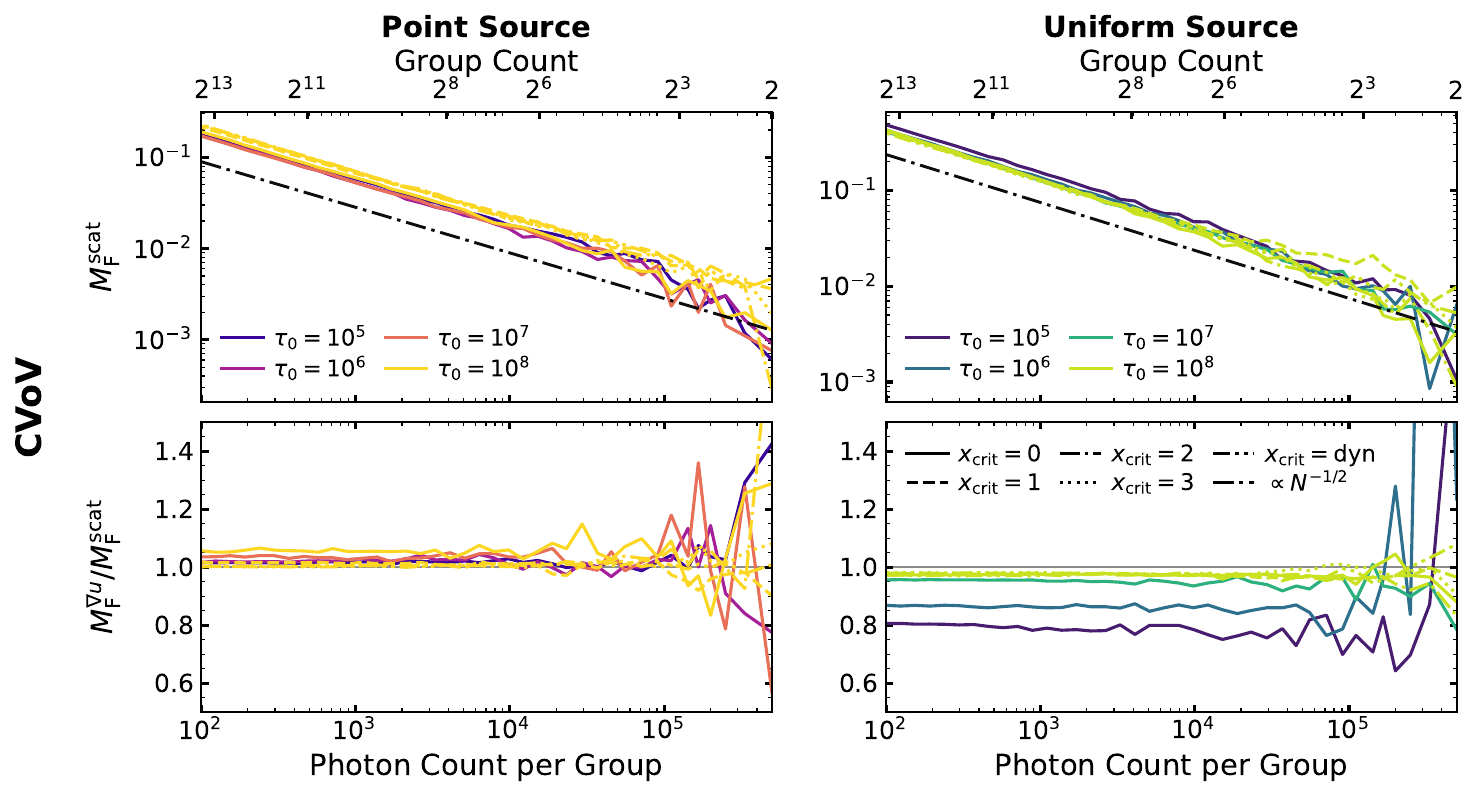}
    \caption{Coefficient of variation of variance as a function of photon count per group. Rows show the direct scattering estimator and the gradient-of-energy-density reconstruction relative to the scattering estimator; columns show point-source and uniform-source emission. The lower axis gives the photon count per group, while the upper axis gives the corresponding number of independent groups. Colours indicate optical depth, line styles indicate core-skipping prescription, and the black dash-dotted line shows the expected $N^{-1/2}$ scaling. CVoV measures the stability of the estimated variance itself. Values below unity indicate that the variance estimate does not fluctuate by an amount comparable to its own mean, supporting the use of the corresponding FE, $N_\mathrm{needed}$, and $t_\mathrm{needed}$ diagnostics.}
    \label{fig:cvov}
\end{figure*}

Fig.~\ref{fig:cvov} shows the coefficient of variation of variance as a function of photon count per group for the direct event-based estimator and the gradient-of-energy-density reconstruction. The reconstruction is shown relative to the scattering estimator, making it clear where the two behaviours deviate or agree. $\nabla \cdot P$ was omitted because its CVoV was nearly identical to $\nabla u$. Across estimator type, source geometry, optical depth, and core-skipping prescription, the CVoV curves decrease approximately as $N^{-1/2}$. This is expected behaviour for independent packet-level samples with a finite fourth moment. The normalisation is above the ideal Gaussian expectation, indicating non-Gaussian and somewhat heavy-tailed contribution distributions, but the important point is that the variance estimates become steadily more stable with increasing group size.

The CVoV values remain below unity throughout the plotted range and fall to much smaller values at larger group sizes. This means that the variance estimate does not fluctuate by an amount comparable to its own mean for the photon counts used in the main diagnostics. The FE, $N_\mathrm{needed}$, and $t_\mathrm{needed}$ curves are therefore statistically meaningful summaries of the finite-sampling error. The scatter at the largest group sizes should not be over-interpreted, because those points are constructed only from a small number of independent groups.

The similarity of the CVoV behaviour across optical depths is also informative. Although the mean force and bias vary strongly with $\tau_0$ and $x_\mathrm{crit}$, the stability of the variance estimate is primarily governed by the packet-level contribution distribution after the estimator has been fully accumulated. Once the full photon history is treated as the elementary Monte Carlo sample, the variance behaves regularly under grouping. This does not mean that the underlying event sequence is uncorrelated. It simply means the event-level covariance is already contained inside each completed packet contribution. The second-order result therefore supports the use of the packet-level moment hierarchy adopted in the convergence hierarchy.

\subsection{Discussion}

The convergence behaviour of Ly$\alpha$ MCRT momentum-transfer estimators must be understood through the full hierarchy. The zeroth-order results show that the first question is what value an estimator is converging toward and whether this is physically sound. For the direct scattering estimator, the no-core-skipping calculation provides the cleanest reference because it retains the full sequence of momentum-depositing events. Core skipping systematically lowers the integrated force multiplier, producing negative bias because skipped core scatterings are part of the internal momentum budget rather than computational overhead alone. This bias becomes less severe at high optical depth, where the force is increasingly dominated by rare excursion events, but it remains important for aggressive skipping prescriptions.

The path-based estimators introduce a different zeroth-order limitation. They do not measure momentum deposition directly. Instead, they reconstruct the force from radiation moments and spatial derivatives. For uniform-source emission, this reconstruction behaves comparatively well, with the path-based curves largely tracking the scattering-based bias trends. For point-source emission, however, the central concentration of the radiation field makes the reconstruction more delicate. The gradient of the energy density and the divergence of the radiation pressure can therefore show estimator-level offsets even when no core scatterings are skipped. The resolution-bias test confirms that this is a spatial reconstruction issue. Once the force-relevant gradient is smeared over too coarse a radial cell, the inferred force no longer matches the direct momentum deposition.

The first-order results then show that the statistical precision is a separate issue from physical accuracy. The relative-error curves demonstrate that each estimator can recover its own full-sample mean with the expected $N^{-1/2}$ scaling. Point-source calculations require only modest photon counts to reach percent-level precision, while uniform-source calculations require roughly an order of magnitude more photons because unbiased volume-weighted launching under samples the geometric core. The FE and $N_\mathrm{needed}$ diagnostics therefore quantify the cost of estimating a given estimator's own target value, not whether the target value is unbiased.

A key result is that the three force constructions have comparable order of magnitude in fractional error. This is not what one would expect from the raw, uncorrelated contribution variances alone. The direct scattering estimator should appear highly noisy at the level of individual kicks, while the path-based estimators should appear much smoother at the level of path-length moments. However, the final estimator variance depends on the covariance structure of the completed photon histories. For the event-based estimator, covariance suppresses much of the raw scattering variance. For the path-based estimators, covariance reinforces the variance of the reconstructed force. The result is that the total FE of the event-based and path-based force estimators lies in the same broad range, with little optical-depth dependence, consistent with the scaling arguments developed earlier.

The runtime analysis adds the final practical layer. Similar values of $N_\mathrm{needed}$ do not imply similar computational cost, because the cost per photon depends strongly on optical depth and core-skipping prescription. The no- and weak-core-skipping cases give the most physically faithful momentum deposition. They also have the steepest runtime scaling with optical depth. Aggressive core skipping and dynamical core skipping reduce the computational cost substantially, with the dynamical prescription showing much weaker optical-depth dependence. This efficiency gain comes at the cost of zeroth-order bias. Thus, core skipping should be understood as a controlled tradeoff between computational speed and physical fidelity. Core skipping is not an algorithm for efficiently reducing variance.

The second-order results validate the use of the first-order diagnostics. The CVoV curves decrease approximately as $N^{-1/2}$ and remain below unity across the sampled cases. This means that the variance estimates are themselves stable, supporting the packet-level convergence hierarchy adopted here. The individual scattering events within a photon history are highly correlated, but once the full photon history is treated as the elementary MC sample, the variance behaves regularly under grouping.

Overall, convergence of Ly$\alpha$ MCRT radiation forces cannot be summarised by a single error curve. Zeroth-order convergence determines whether the estimator is targeting the correct physical force. First-order convergence determines how many photons are needed to estimate that target with a desired precision. Runtime analysis determines how expensive those photons are. Second-order convergence determines whether the quoted variance is itself reliable. The direct scattering estimator without core skipping is the most trustworthy reference for momentum transfer, while path-based estimators and core-skipped calculations can be useful and efficient only when their reconstruction bias, sampling error, and computational tradeoffs are assessed together.

\section{Summary and conclusions}
\label{sec:conclusions}

In this work, we have treated the convergence of internal Ly$\alpha$ radiation-force calculations as an estimator problem. Rather than asking whether a Monte Carlo calculation appears smooth, or whether a sufficiently large number of photons has been used, we developed a hierarchy that separates convergence into three tiers. This distinction is essential for resonant-line transport because internal quantities depend on both photon histories inside the gas and the properties of photons that eventually escape. The main conclusions are as follows:

\begin{enumerate}
\item Convergence of internal Ly$\alpha$ forces must be addressed at the level of the estimator distribution. A force calculation should not be declared fully converged unless every order of the hierarchy has also converged. Apparent convergence in one diagnostic is not sufficient to establish total convergence, since each diagnostic probes a different aspect of the force calculation. Failure at any level can make a seemingly stable internal Ly$\alpha$ force estimate misleading.

\item The convergence hierarchy also clarifies the role of estimator choice. The direct scattering estimator most closely represents the physical momentum exchanged between photons and gas, but it becomes computationally expensive when all resonant scatterings are retained. Path-based force reconstructions offer a complementary approach by estimating forces through radiation-field moments. However, their accuracy depends on the validity of the reconstruction procedure and on sufficient spatial resolution.

\item Core-skipping algorithms reduce computational cost, but in the context of internal force calculations they should be interpreted as modifying the target momentum budget, not merely as accelerating numerical convergence.

\item Consequently, direct scattering estimators, core-skipped calculations, and path-based estimators should not be regarded as interchangeable numerical routes to the same result. Rather, they exhibit different tradeoffs between bias, variance, and computational cost.

\item The main conclusion is that reliable internal Ly$\alpha$ force calculations require simultaneous attention to the various orders of convergence through estimator bias, variance, stability, and computational cost.

\end{enumerate}



The framework developed here is therefore intended as a practical diagnostic tool for future Ly$\alpha$ MCRT calculations. For any internal force calculation, one should ask whether the estimator has been benchmarked against an appropriate mean, whether its finite-sampling error is below a designated threshold, and whether the estimated variance is stable under resampling. These checks are especially important when acceleration schemes are used, because reducing runtime or apparent noise does not guarantee that the physical force has been preserved.

Several extensions follow naturally from this work. The first is a more detailed study of the covariance structure of the momentum-transfer estimator. The present paper uses packet-level moments so that correlations within each photon history are automatically included in the measured variance. It does not attempt to identify where those correlations arise or how they depend on frequency, spatial region, or scattering physics. A companion study will decompose the estimator into physically meaningful pieces in order to determine where covariance suppresses or reinforces the final variance. It will also address to what extent this behaviour is specific to Ly$\alpha$ transport rather than a more general feature of scattering Monte Carlo.

A second direction is to turn the convergence hierarchy into an optimisation principle. Once the relevant error metric is defined, sampling strategies can be designed to minimise estimator variance at fixed computational cost rather than simply to increase photon counts globally. This motivates adaptive importance-sampling schemes in which photon launching, weighting, and stopping criteria are adjusted using local estimates of contribution and noise. Such methods are not limited to Ly$\alpha$ radiation forces. The same logic applies to any Monte Carlo estimator whose accuracy depends on the distribution of weighted packet contributions.

The broader lesson is that convergence should be treated as part of the estimator design. An after-the-fact visual check is not sufficient. Reliable MCRT requires separating accuracy from precision and stability. The moment-based hierarchy introduced here provides a systematic way to make that separation, and a foundation for designing future estimators that are both faster and more trustworthy.

\section*{Acknowledgements}
We thank Shyam Menon and Shayaan Zari for discussions that benefited this work. All computations were performed on the Juno cluster supported by UT Dallas. AS acknowledges support through HST AR-17859, HST AR-17559, and JWST AR-08709.



\bibliographystyle{mnras}
\bibliography{main}




\begin{appendix}
\renewcommand{\thefigure}{\thesection\arabic{figure}}
\renewcommand{\thetable}{\thesection\arabic{table}}

\section{Spatial-derivative stencils for path-based force reconstructions}
\label{app:spatial_stencils}

The path-based force estimators used in the main text require a spatial derivative of Monte Carlo radiation-field moments. This appendix summarises the two derivative stencils used for those reconstructions. The purpose is to separate the Monte Carlo estimator for the underlying radiation moment from the deterministic operation that converts that moment into an acceleration. This distinction is important because the resolution tests in Fig.~\ref{fig:res-bias} probe the latter: they test how accurately a finite radial grid can reconstruct a force from cell-centred estimates of $u$ or $P_{rr}$.

We denote the radial cell faces by $r_{i-1/2}$ and $r_{i+1/2}$, the cell centre by $r_i$, the shell volume by
\begin{equation}
\Delta V_i =
\frac{4\pi}{3}
\left(
r_{i+1/2}^3-r_{i-1/2}^3
\right) \, ,
\end{equation}
and the spherical face areas by
\begin{equation}
A_{i\pm 1/2}=4\pi\,r_{i\pm 1/2}^2 \, .
\end{equation}
The Monte Carlo path-length estimators provide cell-centred values of the radiation energy density $u_i$ and radial radiation-pressure component $P_{rr,i}$. The derivative stencil then maps these cell-centred moments into a cell-centred acceleration.

\subsection{Centred finite-difference stencil}

The simplest reconstruction differentiates the cell-centred radiation moment directly. For the gradient-of-energy-density estimator, the interior-cell finite-difference stencil is
\begin{equation}
\left(\frac{\text{d}u}{\text{d}r}\right)_i^{\rm FD}
=
\frac{u_{i+1}-u_{i-1}}
{r_{i+1}-r_{i-1}} \, ,
\end{equation}
with one-sided differences used at the radial boundaries. The corresponding acceleration is
\begin{equation}
a_i^{\nabla u,{\rm FD}}
=
-\frac{1}{3\rho_i}
\left(\frac{\text{d}u}{\text{d}r}\right)_i^{\rm FD} \, .
\end{equation}
This stencil is straightforward and becomes equivalent to the finite-volume reconstruction at sufficiently high resolution, but it does not explicitly enforce the shell-integrated conservation law. It can therefore be sensitive to steep central gradients, boundary conditions, and coarse radial binning.

For the radiation-pressure reconstruction, the radial component of the divergence of a spherically symmetric radiation-pressure tensor is
\begin{equation}
\left(\nabla\cdot \mathbf{P}\right)_r
=
\frac{\text{d}P_{rr}}{\text{d}r}
+
\frac{3P_{rr}-u}{r} \, ,
\end{equation}
where we used the trace relation
\begin{equation}
P_{\theta\theta}+P_{\phi\phi}=u-P_{rr} \, .
\end{equation}
The finite-difference version therefore uses
\begin{equation}
\left(\nabla\cdot \mathbf{P}\right)_{r,i}^{\rm FD}
=
\frac{P_{rr,i+1}-P_{rr,i-1}}
{r_{i+1}-r_{i-1}}
+
\frac{3P_{rr,i}-u_i}{r_i} \, ,
\end{equation}
and
\begin{equation}
a_i^{\nabla\cdot P,{\rm FD}}
=
-\frac{1}{\rho_i}
\left(\nabla\cdot \mathbf{P}\right)_{r,i}^{\rm FD} \, .
\end{equation}
The second term is the spherical-geometry curvature term. It vanishes in the isotropic diffusion limit, where $P_{rr}=u/3$, leaving the expected relation $(\nabla\cdot\mathbf{P})_r=(1/3) \, \text{d}u/\text{d}r$.

\subsection{Finite-volume stencil}

The second reconstruction is a finite-volume stencil. Instead of differentiating the cell-centred profile directly, it estimates the shell-averaged force from the flux of radiation pressure through the inner and outer spherical faces. Integrating the radial pressure divergence over shell $i$ gives
\begin{equation}
\left\langle
\left(\nabla\cdot \mathbf{P}\right)_r
\right\rangle_i^{\rm FV}
=
\frac{1}{\Delta V_i}
\left[
A_{i+1/2}P_{rr,i+1/2}
-
A_{i-1/2}P_{rr,i-1/2}
-
C_i
\right] \, ,
\end{equation}
where $P_{rr,i\pm1/2}$ are face-centred pressure estimates and
\begin{equation}
C_i
=
\int_{r_{i-1/2}}^{r_{i+1/2}}
4\pi r
\left(
P_{\theta\theta}+P_{\phi\phi}
\right)
\text{d}r
\end{equation}
is the spherical-geometry correction. Using
$P_{\theta\theta}+P_{\phi\phi}=u-P_{rr}$ and taking the cell-centred moments to be constant within the shell gives
\begin{equation}
C_i
\approx
2\pi
\left(
r_{i+1/2}^2-r_{i-1/2}^2
\right)
\left(
u_i-P_{rr,i}
\right) \, .
\end{equation}
The reconstructed acceleration is then
\begin{equation}
a_i^{\nabla\cdot P,{\rm FV}}
=
-\frac{1}{\rho_i}
\left\langle
\left(\nabla\cdot \mathbf{P}\right)_r
\right\rangle_i^{\rm FV} \, .
\end{equation}

For the gradient-of-energy-density estimator, the same finite-volume formula is obtained by taking the diffusion-limit closure $P_{rr}=u/3$. In that case,
\begin{equation}
C_i^{\nabla u}
\approx
\frac{4\pi}{3}
u_i
\left(
r_{i+1/2}^2-r_{i-1/2}^2
\right) \, ,
\end{equation}
and
\begin{equation}
a_i^{\nabla u,{\rm FV}}
=
-\frac{1}{\rho_i \Delta V_i}
\left[
A_{i+1/2}\frac{u_{i+1/2}}{3}
-
A_{i-1/2}\frac{u_{i-1/2}}{3}
-
C_i^{\nabla u}
\right] \, .
\end{equation}
Interior face values are reconstructed by linear interpolation from neighboring cell centres,
\begin{equation}
X_{i+1/2}=\frac{1}{2}\left(X_i+X_{i+1}\right) \, ,
\end{equation}
for $X=u$ or $P_{rr}$. The inner boundary is regular because $A_{1/2}=0$ at the origin. At the outer boundary, the default reconstruction uses a vacuum-like closure for the outgoing face pressure.

The finite-volume stencil has two useful properties. First, it respects the shell-integrated divergence law, so the reconstructed acceleration is tied directly to the net pressure force on the cell. Second, it treats the geometric terms in spherical coordinates explicitly. For example, a constant isotropic radiation pressure should not produce a force; the finite-volume geometry correction cancels the apparent face-area imbalance in this limit. The centred finite-difference stencil does not enforce this cancellation at the shell-integrated level, which makes it more vulnerable to coarse radial grids and steep point-source gradients.

Both stencils converge to the same continuum derivative as the radial resolution is increased. The differences seen in Fig.~\ref{fig:res-bias} should therefore be interpreted as finite-resolution reconstruction effects, not as differences in the underlying Monte Carlo path-length tallies. The same photon histories provide the estimates of $u_i$ and $P_{rr,i}$; only the deterministic spatial derivative used to convert those moments into a force has been changed.

\section{Volume-boosted uniform-source tests}
\label{app:uniform_volume_boosting}

\begin{figure}
    \centering
    \includegraphics[width=\linewidth]{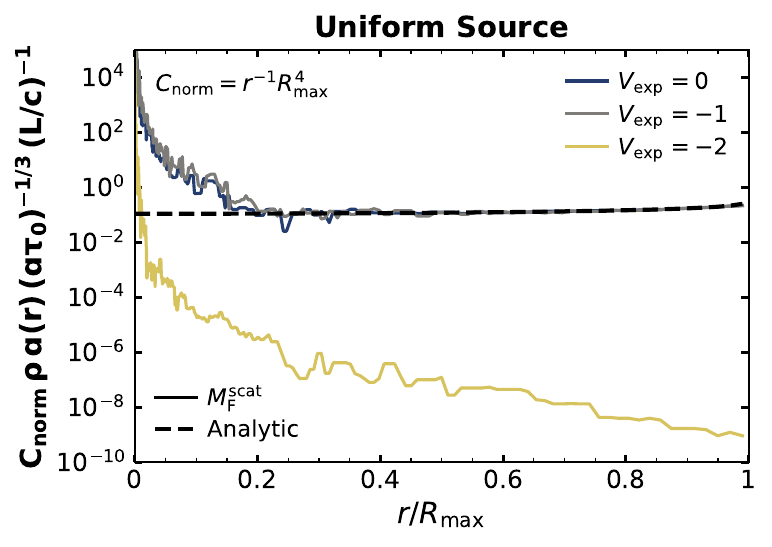}
    \caption{
    Uniform-source acceleration profiles for volume-boosted launching. Curves compare unbiased launching, $V_\mathrm{exp}=0$, against boosted cases with $V_\mathrm{exp}<0$. Moderate boosting theoretically extends the usable profile farther into the geometric core by increasing the number of centrally launched photons, but it does not substantially repair the broader core-region mismatch with the analytic profile. Stronger boosting can produce a systematic downward shift at fixed photon count, indicating that the weighted estimator has not reached a stable finite-sample mean.
    }
    \label{accel_uniform_profile_boost}
\end{figure}

The main text uses unbiased uniform-source launching, because this is the direct Monte Carlo sampling of the physical emissivity distribution. In a uniform sphere, however, unbiased launching samples photon birth positions in proportion to volume. This is correct in expectation, but it leaves the small-volume central region with relatively few photon histories at finite photon count. Since the uniform-source acceleration profile is most difficult to measure in this geometric core, we use volume-boosted runs as a supplementary test of whether increased central sampling improves the convergence diagnostics.

These tests should be interpreted as importance-sampling diagnostics rather than as a separate physical model. Volume boosting changes the launch distribution but applies compensating packet weights so that the target emissivity is preserved in expectation. Moderate boosting can increase the number of photon histories that probe the inner cloud, but stronger boosting also increases the dispersion of packet weights. The result is a tradeoff between improved core sampling and increased weight-driven variance. Therefore, the boosted runs do not replace the unbiased results used in the main text; they test whether the uniform-source core behaviour is primarily a sampling problem and whether simple launch biasing is an effective remedy.

\subsection{Launch biasing method}

For uniform sources, photon packets are generated according to the luminosity assigned to each cell,
\begin{equation}
\mathcal{L}_i
=
\int_{V_i}
\epsilon(\bm{r})\,\text{d}V \, ,
\end{equation}
where $\epsilon$ is the local emissivity. In the uniform case, $\epsilon$ is constant and therefore $\mathcal{L}_i\propto V_i$. The unbiased launch probability is
\begin{equation}
p_i =
\frac{\mathcal{L}_i}{\sum_j \mathcal{L}_j} \, ,
\end{equation}
so $p_i\propto V_i$ for equal-emissivity cells.

The volume-boosted launch distribution modifies this probability by an additional volume-dependent factor,
\begin{equation}
q_i =
\frac{\mathcal{L}_i (V_i/V_{\rm tot})^{\eta_V}}
{\sum_j \mathcal{L}_j (V_j/V_{\rm tot})^{\eta_V}} \, ,
\end{equation}
where $\eta_V=0$ gives unbiased launching and $\eta_V<0$ boosts smaller-volume cells. In the notation used in the figures, $\eta_V=V_{\rm exp}$. Each photon launched from cell $i$ receives a compensating weight
\begin{equation}
w_i =
\frac{p_i}{N_{\rm ph} q_i} \, ,
\end{equation}
so the weighted estimator preserves the original uniform-source luminosity distribution in expectation.

At finite photon count, however, the reweighting changes the distribution of packet contributions. Regions oversampled by $q_i$ receive smaller weights, while regions undersampled by $q_i$ receive larger weights. Moderate boosting can reduce core undersampling, but aggressive boosting can broaden the weight distribution enough to destabilise the finite-sample mean or increase FE. The boosted tests below therefore use the same bias, FE, and CVoV diagnostics as the main text.

\subsection{Volume-boosted tests}
\label{app:uniform_accel_boosting}

\begin{figure}
    \centering
    \includegraphics[width=\linewidth]{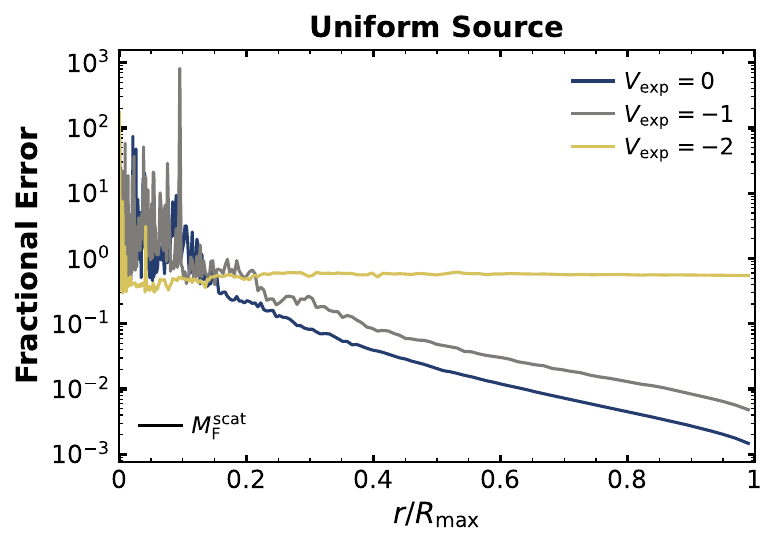}
    \caption{
    Cumulative fractional error for the volume-boosted uniform-source momentum-transfer estimator. Boosting redistributes the noise rather than providing a clean variance-reduction win. Moderate boosting does not significantly suppress the large core fluctuations, while stronger boosting can flatten some inner spikes at the cost of raising the cumulative FE through much of the domain.
    }
    \label{accel_uniform_fracerror_boost}
\end{figure}

Several diagnostics are tested with volume boosting, primarily the radial acceleration profile, fractional error, bias, and CVoV. The goal is to determine whether volume boosting is a viable tool for reducing variance or repairing the undersampling issue found in uniform-source tests. The different force reconstructions have been omitted because they follow the same trends as the direct scattering estimator.

\subsubsection{Boosted acceleration profile}

Fig.~\ref{accel_uniform_profile_boost} displays the volume-boosted acceleration profile. The comparison shows that $V_\mathrm{exp}=-1$ has only a limited impact on the overall profile shape. It primarily provides more usable sampling at extremely small radii, extending the profile inward, but it does not substantially repair the broader core-region mismatch with the analytic trend.

In contrast, $V_\mathrm{exp}=-2$ changes the profile dramatically and in an unfavorable way. The boosted curve is pulled downward, crosses the analytic expectation at small radius, and then remains systematically below the analytic trend through much of the domain. This indicates a strong underestimation of the acceleration profile across a wide radial range at this photon count.

Since importance sampling should preserve unbiased means in the infinite-sample limit, this behaviour should be interpreted as a finite-sample instability of the weighted estimator. At $N_\mathrm{ph}=10^6$, aggressive boosting appears to produce enough weight dispersion that the sample mean is dominated by a small number of effectively influential histories.

\subsubsection{Boosted cumulative fractional error}

The FE comparison reinforces this picture. In Fig.~\ref{accel_uniform_fracerror_boost}, $V_\mathrm{exp}=-1$ does not meaningfully suppress the large, spiky core noise, and it tends to raise the cumulative FE in the bulk of the domain relative to $V_\mathrm{exp}=0$. The $V_\mathrm{exp}=-2$ case flattens some of the largest inner spikes at the expense of a higher cumulative FE through much of the geometry. This again implies a higher overall variance cost at fixed $N_\mathrm{ph}$.

Thus, for the momentum-transfer estimator, volume boosting is not a clean variance-reduction scheme. Boosting increases core sampling and also introduces broader weight dispersion. Since radiation force already has a heavy-tailed contribution distribution, the net result is that boosting does not necessarily reduce the total variance and can instead increase it.

\subsubsection{Boosted integrated bias}

\begin{figure}
    \centering
    \includegraphics[width=\linewidth]{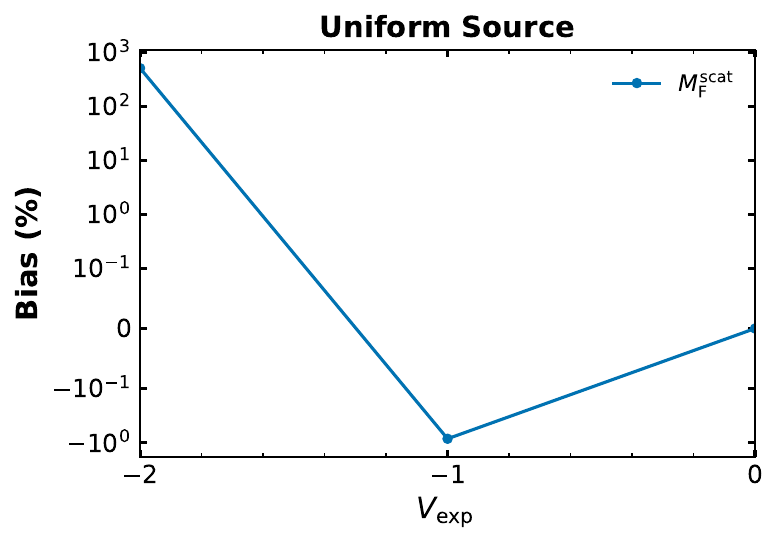}
    \caption{
    Integrated force-multiplier bias for the uniform-source scattering-based construction, including volume-boosted launch distributions. Moderate boosting remains broadly consistent with the analytic benchmark, while aggressive boosting produces a large positive finite-sample bias. This indicates that strong boosting can destabilise the weighted acceleration estimator at the tested photon count.
    }
    \label{accel_uniform_bias_boost}
\end{figure}

The boosted integrated-bias comparison in Fig.~\ref{accel_uniform_bias_boost} makes the practical outcome clear. The integrated $M_\mathrm{F}$ bias remains essentially consistent with the analytic benchmark for $V_\mathrm{exp}=0$ and $V_\mathrm{exp}=-1$. However, $V_\mathrm{exp}=-2$ produces an extremely large positive bias, far larger than the shifts induced by core skipping in the unbiased $x_\mathrm{crit}$ sweep.

This again points to the same conclusion: at the tested photon count, aggressive boosting does not behave as a stable variance-reduction tool for the momentum-transfer estimator. Instead, it appears to push the weighted estimator into a regime where the mean is dominated by a small number of effectively influential histories.

\subsubsection{Boosted CVoV}

\begin{figure}
    \centering
    \includegraphics[width=\linewidth]{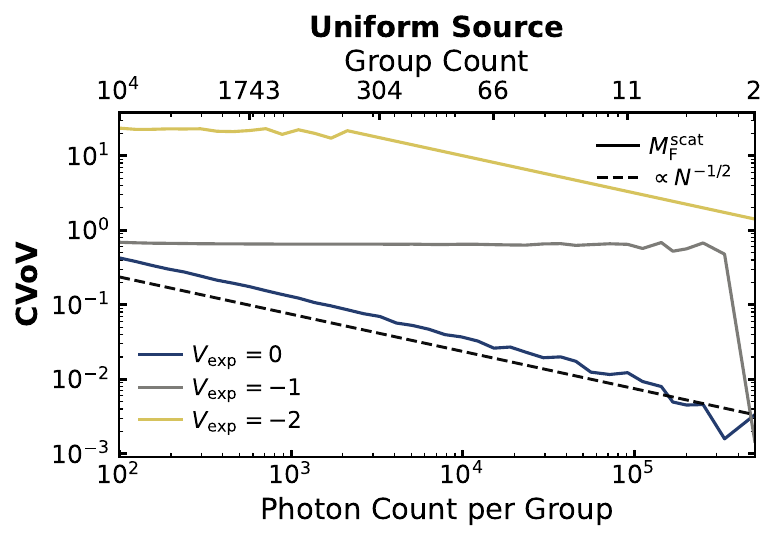}
    \caption{
    CVoV for volume-boosted uniform-source calculations. Boosting violates the expected $N^{-1/2}$ group-size scaling, and it changes the normalisation by modifying the tail structure of the per-photon contribution distribution.}
    \label{accel_uniform_cvov_boost}
\end{figure}

Fig.~\ref{accel_uniform_cvov_boost} shows that volume boosting alters the fundamental CVoV scaling with group size. Even moderate volume boosting raises CVoV to order unity and weakens the expected scaling. For stronger volume boosting, CVoV exceeds order unity. It is difficult to determine whether the low-group-count trends are due to finite sampling or to changes in the weighted contribution distribution. Regardless, the outcome is clear. Volume boosting increases the variance of the variance, making the lower-order convergence metrics less trustworthy.

\subsection{Summary}

Taken together, these tests show that volume boosting is not an effective remedy for the uniform-source undersampling problem. Moderate boosting, $V_{\rm exp}=-1$, does not repair the broader disagreement with the analytic solution and provides no consistent reduction in fractional error. Strong boosting, $V_{\rm exp}=-2$, performs substantially worse, producing a distorted acceleration profile, large integrated bias, and greater variance instability. Although the compensating weights preserve the correct expectation value asymptotically, they broaden the distribution of per-photon contributions and reduce the effective sample size at finite $N_{\rm ph}$. This is reflected by the elevated CVoV and delayed recovery of the expected $N^{-1/2}$ scaling. Consequently, the gain in central photon sampling is offset, or exceeded, by weight-driven noise, and volume boosting is not adopted as a variance-reduction strategy.

\end{appendix}

\end{document}